\documentclass[epj]{svjour}

\usepackage{color}
\usepackage{xcolor}
\definecolor{bred}{rgb}{0.8, 0.0, 0.0}
\definecolor{pblue}{rgb}{0.2, 0.2, 0.6}
\definecolor{ao}{rgb}{0.0, 0.5, 0.0}
\definecolor{carmine}{rgb}{0.59, 0.0, 0.09}

\newcommand{\keVee}{\,keV$_{ee}$}

\newcommand{\CEvNS}{CE$\nu$NS}

\usepackage{graphicx}
\usepackage{amsmath}
\usepackage{amssymb}
\usepackage{hyperref}
\usepackage{cite}
\usepackage{arydshln}
\usepackage{upgreek}
\usepackage{subcaption}

\usepackage{hyperref}
\hypersetup{
    colorlinks = true,
    linkbordercolor = {white},
    linkcolor = bred,
    citecolor = pblue,
    urlcolor = pblue
}

\usepackage[switch]{lineno} 

\begin{document}\sloppy

\title{Pulse shape discrimination for the CONUS experiment in the keV and sub-keV regime}

\titlerunning{Pulse Shape Discrimination for the \textsc{Conus} experiment in the keV and sub-keV regime}
\author{H.~Bonet\inst{1}, A.~Bonhomme\inst{1}, C.~Buck\inst{1}, K. F\"{u}lber\inst{2}, J. Hakenm\"{u}ller\inst{1}, J.~Hempfling\inst{1}, J.~Henrichs\inst{1}, G.~Heusser\inst{1}, M.~Lindner\inst{1}, W.~Maneschg\inst{1}, T.~Rink\inst{1}, E.~S\'{a}nchez Garc\'{i}a\inst{1}, J.~Stauber\inst{1}, H.~Strecker\inst{1}, R.~Wink\inst{2}
}
\authorrunning{H. Bonet et al.}
\institute{Max-Planck-Institut f\"ur Kernphysik, Saupfercheckweg 1, 69117 Heidelberg, Germany \and PreussenElektra GmbH, Osterende, Brokdorf, Germany \vspace*{0.2cm}\\ E-mail address: \href{mailto:conus.eb@mpi-hd.mpg.de}{conus.eb@mpi-hd.mpg.de}}

\date{}

\abstract{
Point-contact p-type high-purity germanium detectors (PPC HPGe) are particularly suited for detection of sub-keV nuclear recoils from coherent elastic scattering of neutrinos or light dark matter particles. While these particles are expected to interact homogeneously in the entire detector volume, specific classes of external background radiation preferably deposit their energy close to the semi-active detector surface, in which diffusion processes dominate that subsequently lead to slower rising pulses compared to the ones from the fully active bulk volume. Dedicated studies of their shape are therefore highly beneficial for the understanding and the rejection of these unwanted events.
This article reports about the development of a data-driven pulse shape discrimination (PSD) method for the four 1\,kg size PPC HPGe detectors of the \textsc{Conus} experiment in the keV and sub-keV regime down to 210\,eV$_{ee}$. The impact of the electronic noise at such low energies is carefully examined. 
It is shown that for an acceptance of 90\,\% of the faster signal-like pulses from the bulk volume, approx. 50\,\% of the surface events can be rejected at the energy threshold and that their contribution is fully suppressed above 800\,eV$_{ee}$. Applied to the \textsc{Conus} background data, such a PSD rejection cut allows to achieve an overall $(15-25)\,\%$ reduction of the total background budget.
The new method allows to improve the sensitivity of future \textsc{Conus} analyses and to refine the corresponding background model in the sub-keV energy region. 
}


\maketitle

\section{Introduction} \label{section:introduction} 

The development of sub-keV energy threshold detectors opened the possibility to detect tiny nuclear recoils from coherent elastic neutrino-nucleus scattering (\CEvNS) or to search for light dark matter interaction. High-purity germanium (HPGe) detectors are specifically well suited for these searches and widely used \cite{Soma2016, Aalseth2013, Collar2021b, Belov2015, Agnolet2017, Augier2021, Arnaud2018, Yang2018}. The \textsc{Conus} experiment aims at detecting the \CEvNS~signal from reactor antineutrino \CEvNS~at the Brokdorf nuclear power plant (Germany) by deploying four 1\,kg size p-type point contact (PPC) germanium detectors, specifically designed to have an ultra-low noise and a very low intrinsic background contamination\cite{Bonet2021d}. Thanks to the very low background level achieved and an energy threshold of 210\,eV$_{ee}$ (electron equivalent energy), \textsc{Conus} has already provided stringent constraints on the Standard Model \CEvNS~interaction\cite{Bonet2021} and neutrino physics beyond it \cite{Bonet2021b, Bonet2022}. The challenge of the detection of \CEvNS~at a reactor site was recently underlined by our precise measurement of the quenching factor in germanium in the sub-keV energy region \cite{Bonhomme2022} and calls for further background reduction efforts and deeper understanding of the signals at energies below 500\,eV$_{ee}$, i.e.\ in the region of interest for the \CEvNS~events search. For the \textsc{Conus} experiment, the background was thoroughly studied and supported by detailed Monte-Carlo (MC) simulation \cite{Bonet2021_bkg}.
It was shown that about half of the background consists of internal contamination inside the end cap that produces radiation in the vicinity of the diode susceptible to interact at the surface of the detectors. In PPC HPGe detectors, these surface interactions lead to anomalous signals, which can fall in the region of interest and complicate data interpretation.
A further step in the background rejection was therefore initiated by recording the pulse waveforms of the signals in order to further discriminate surface from bulk events.

This article reports the first results on pulse shape studies performed on \textsc{Conus} data and it is organized as follows. The origin of the different types of pulse shapes is exposed in Sec.\,\ref{section:physics}. The methods developed to study and categorize the different pulses are detailed in Sec.\,\ref{section:methods}. Their validation and the improvements expected for the \textsc{Conus} data analyses are discussed in Sec.\,\ref{section:results} and is followed by the conclusion in Sec.\,\ref{section:conclusion}.

\section{Signal formation, readout and processing} \label{section:physics}

\subsection{Origin of pulses} \label{subsection:origin_pulses}

\begin{figure*}
		\centering
		\begin{subfigure}[c]{0.22\textwidth}
			\includegraphics[width=\textwidth]{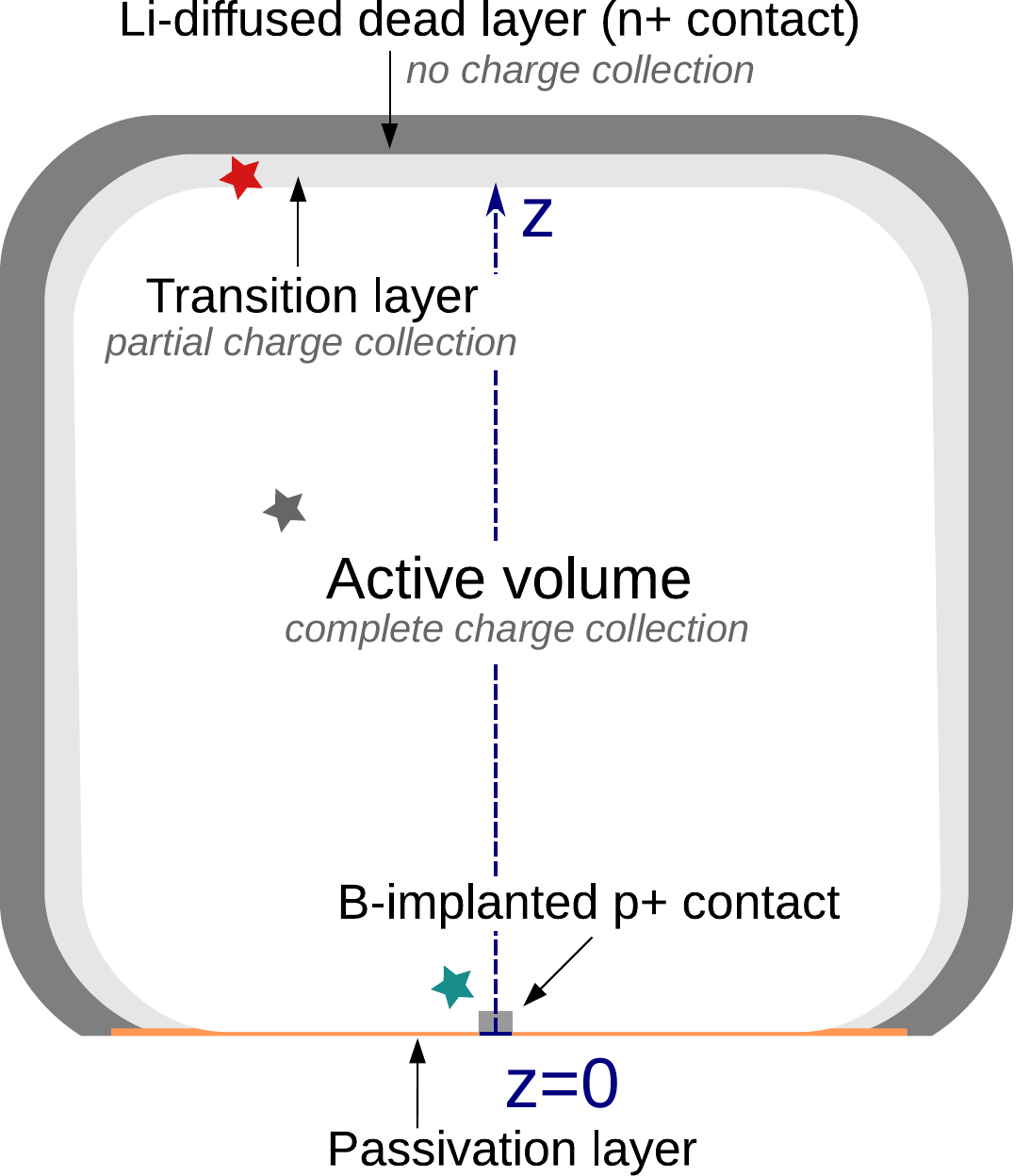}
			\caption{}\label{fig:scheme_diode}
		\end{subfigure}
		\begin{subfigure}[c]{0.33\textwidth}
			\includegraphics[width=\textwidth]{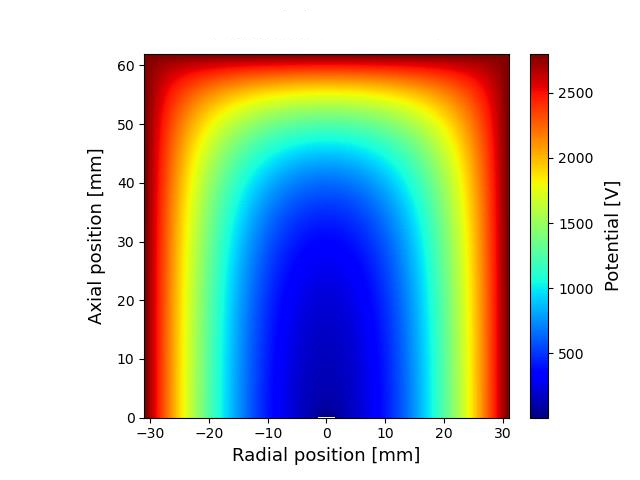}
			\caption{}\label{fig:field_diode}
		\end{subfigure}
		    \begin{subfigure}[c]{0.33\textwidth}
			\includegraphics[width=\textwidth]{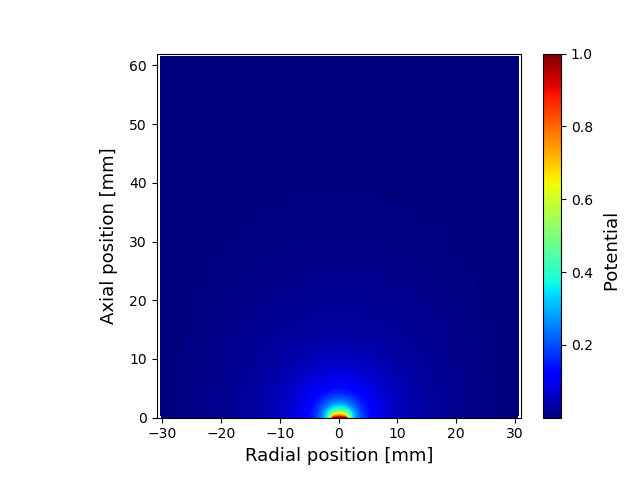}
			\caption{}\label{fig:weighting_potential_diode}
		\end{subfigure}
		\caption{Scheme of a diode used for the \textsc{Conus} detectors (a) with its simulated electric potential (b) and weighting potential (c).}\label{fig:diode}
\end{figure*}

\begin{figure*}
		\centering
		\begin{subfigure}[c]{0.45\textwidth}
			\includegraphics[width=\textwidth]{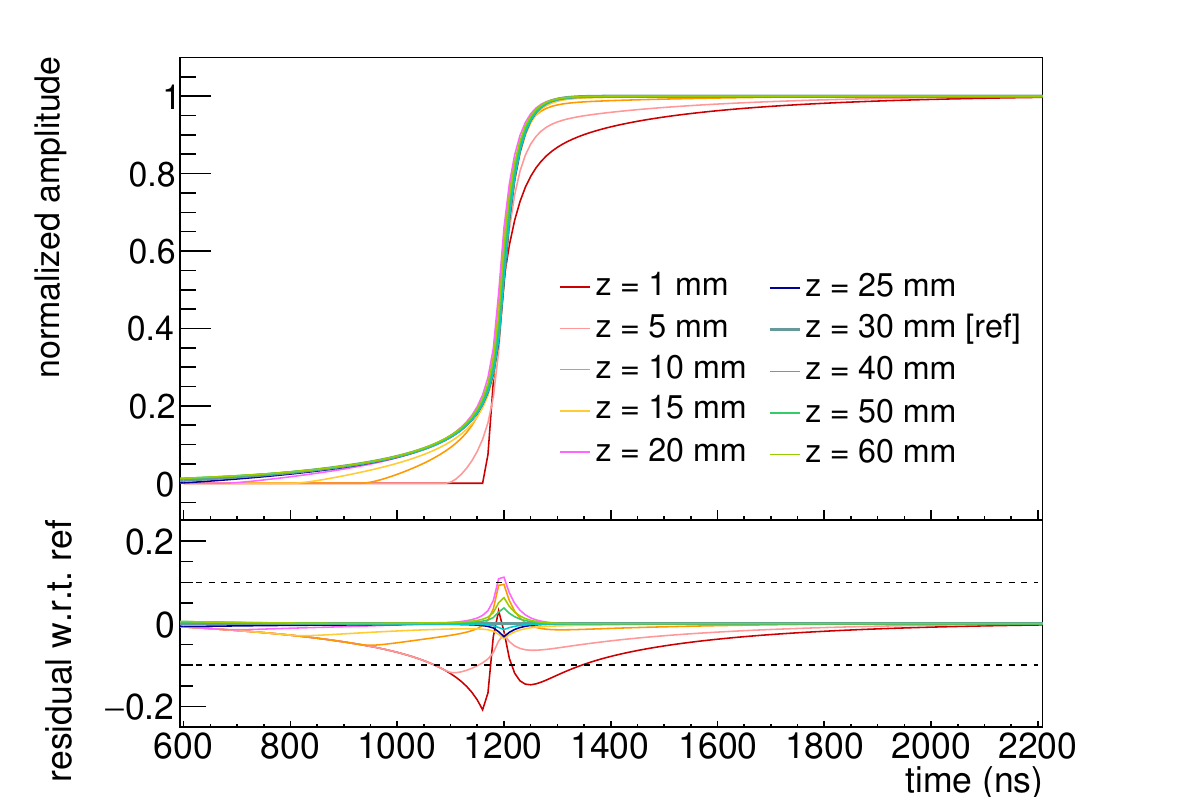}
            \caption{}
            \label{fig:simulated_raw_pulses}
		\end{subfigure}
		\begin{subfigure}[c]{0.45\textwidth}
			\includegraphics[width=\textwidth]{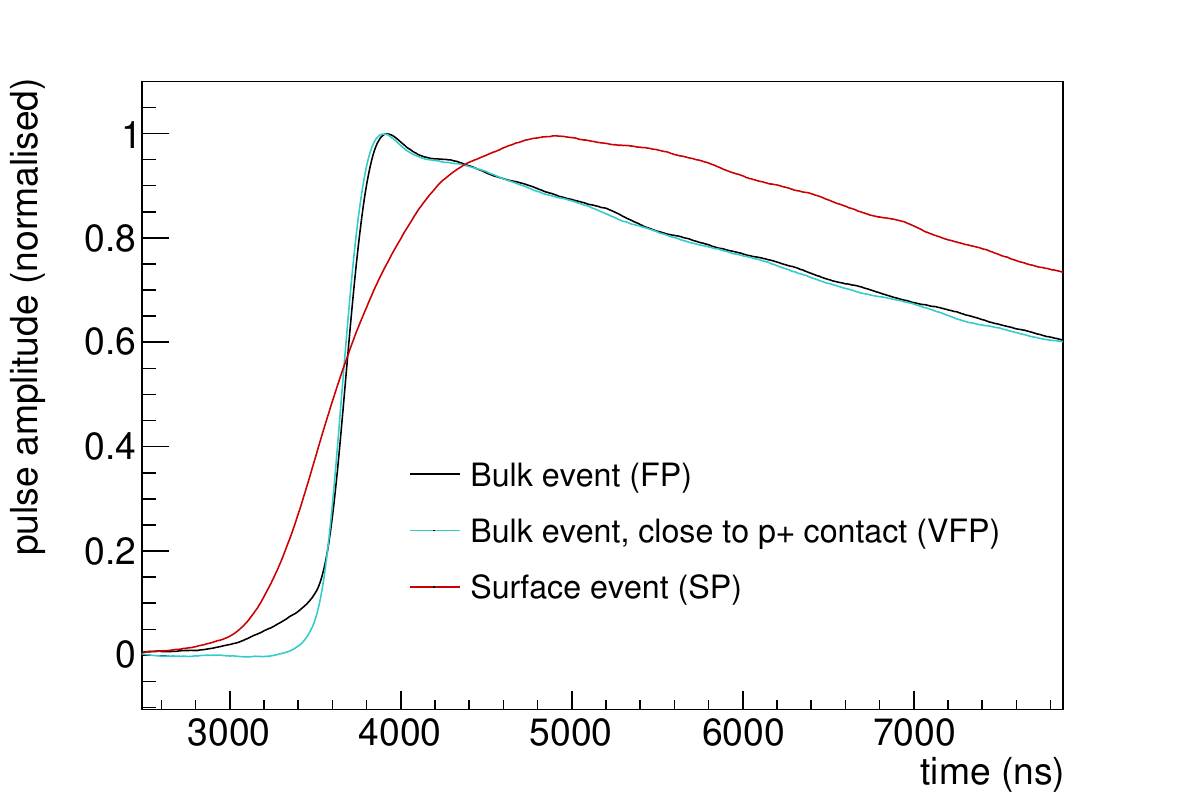}
            \caption{}
            \label{fig:three_pulses}
		\end{subfigure}
		\caption{(a): Simulated charge signal $Q(t)$ induced at the p+ readout electrode for single energy depositions at different interaction vertices along the vertical axis of the cylindrical diodes with height 62\,mm -- represented in Fig.\,\ref{fig:scheme_diode}. Residuals of the pulses w.r.t. a reference signal in the center of the diode volume (z=30\,mm) are shown in the bottom panel. Deviations between pulses become negligible ($\lesssim$10\%) for interactions vertices farther away than $\sim$\,20\,mm from the p+ contact. (b): Measured pulses for the \textsc{Conus-1} detector (each averaged over ten waveforms of 10\keVee~events) for a generic bulk event interaction (black), a bulk interaction in the vicinity of the p+ contact (blue) and a surface event interaction (red). The decay following the rising edge is due to the response of the electronics (see Sec. \ref{subsection:daq}).}
\end{figure*}

The \textsc{Conus} experiment uses four 1\,kg sized p-type point contact high-purity germanium (PPC HPGe) detectors to pursue the goal of the detection of sub-keV$_{ee}$ signals from \CEvNS. The design of the cylindric Ge diodes (diameter and height of 62\,mm) is shown in Fig.\,\ref{fig:scheme_diode}. A full description of the \textsc{Conus} detectors can be found in \cite{Bonet2021d}.
The detectors are electrically depleted by applying a positive reverse bias voltage of a few 1000\,V on the surrounding n+ contact, fabricated by lithium diffusion. The corresponding layer has a typical thickness of 1\,mm and can be decomposed into a dead layer (DL) and a semi-active transition layer (TL) subject to incomplete charge collection. In particular, the TL is evaluated to be $(0.16-0.24)\,\text{mm}$ thick from calibration measurements\cite{Bonet2021d}. The boron-implanted p+ contact has a thickness of 100-200\,nm and a small diameter of $(2.5-3)\,\text{mm}$; it guarantees a detector capacitance $\lesssim$\,1\,pF, which is crucial for low-noise applications. The bottom layer separating p+ and n+ contacts is fully passivated.
Ionizing events in the electrically fully depleted bulk volume\footnote{In the case of the \textsc{Conus} detectors, it corresponds to 91-95\% of the total crystal volume \cite{Bonet2021d}.} create electron-hole pairs  proportional to the deposited energy which drift in opposite directions according to the electric field created by both the applied positive bias voltage and the space charge of the depleted HPGe diode.
The characteristic associated fast pulses (FP) have a rise time of about 300\,ns.
However, an energy deposition within the TL experiences a very weak electric field and diffuses until it either recombines or reaches the bulk volume. This leads to partial charge collection and a larger collection time due to the diffusion process (typically 1\,$\mu$s). Such events will hence be referred thereafter as slow pulses (SP). These events have an erroneous energy reconstruction which especially enhances the event rate at energies below 10\,keV. 
While particles of interest such neutrinos or dark matter particles interact in the entire detector volume, i.e. surface and bulk, and lead therefore mainly to FPs, specific classes of background radiation are absorbed preferably in the outer detector layers leading to the formation of SPs. This difference can be used, to discriminate background from signal-like events. 

To better understand the formation of pulses inside the bulk volume, the electric potential inside the diode was calculated using \texttt{fieldgen}, a part of the \texttt{siggen} software package\cite{Radford2014, alvis2019multisite}. It is shown for illustration in Fig.\,\ref{fig:field_diode}. The p+ contact -- collecting holes -- is used for the readout. 
The simulation framework \texttt{siggen} for signal pulse formation in germanium detectors consists of two parts: the field generation code \texttt{mjd\char`_fieldgen} and the signal generation code \texttt{mjd\char`_siggen}.
Using the known crystal impurity concentrations as well as their gradients, the dead layer and passivation layer thickness and the size of the point contact as detector specific input parameters, the electric field within the diode is determined.
If for a given radiation source the distribution and location of the energy depositions within the diode are known e.g. from Geant4 MC simulations, the expected read-out signal is determined by \texttt{siggen} according to the physics described in the following.
The time-dependent charge signal $Q(t)$ induced at the readout electrode by the movement of the charge carriers can be calculated using the Shockley-Ramo theorem \cite{Zhong2001}:
\begin{equation}
    Q(t) = -q_0\left[\,W(\mathbf{r}_h(t))-W(\mathbf{r}_e(t))\,\right]\, ,
\end{equation}
where $q_0$ represents the total charge carried by the holes and electrons, $\mathbf{r}_{h/e}(t)$ their trajectory paths, and $W$ the weighting potential, i.e. the electric potential for the readout electrode and the others electrodes at unit and zero potential, respectively. The corresponding dimensionless $W$ for one \textsc{Conus} diode is represented in Fig.\,\ref{fig:weighting_potential_diode}. The induced charge $Q(t)$ increases over time until all charges are collected at the electrodes. Due to the smallness of the p+ contact, the weighting potential is homogeneous over a large part of the diode volume and becomes large only within a few mm around the p+ contact. The induced charge mainly results from the drift of the charges in these high $W$ regions and therefore, the position dependency of the signal is rather weak for interactions within the full volume. Induced signals shapes only start to differentiate for interactions very close to the p+ contact. This is illustrated in Fig.\,\ref{fig:simulated_raw_pulses}, where the induced charge signal $Q(t)$ at the readout electrode was calculated using \texttt{siggen} for different interaction points inside the diode volume.
From these simulations, it appears that pulses start to have \textit{very fast} rise times (very fast pulses, hereafter VFP) for a distance from the point contact smaller than 20\,mm, about ten times the size of the p+ contact. Outside of this volume, bulk events have a very similar shape. 
A pulse corresponding to a vertex far from the p+ contact will then simply be delayed compared to a signal from an interaction closer to it, since the charges first have to drift before entering the high field region where they induce a significant charge signal. This characteristic feature provides a very good sensitivity above $\sim\,1$\keVee~to distinguish single-site interactions from multi-site (MS) interactions, resulting for example from multiple Compton scattering processes, whose energy depositions are clearly distinguishable in their signal shape. Such shape analyses were successfully implemented via e.g. the A/E method with low capacitance large HPGe detectors \cite{Agostini2022}.
Exemplary recorded waveforms from different interaction vertices inside the \textsc{Conus-1} diode are shown in Fig.\ref{fig:three_pulses} for illustration.

\subsection{Data acquisition and processing}\label{subsection:daq}
 
Charge signals at the p+ contact are read out by a custom-built transistor reset preamplifier (TRP), chosen to fulfill both ultra-low noise and low background specifications \cite{Bonet2021d}. In order to tag spurious events following the reset of the baselines after saturation of the dynamic range, inhibit TRP signals are generated. Artificially generated charge signals can be injected at the test input of the preamplifier allowing to monitor the response of the whole electronic chain.

\begin{figure}
    \centering
    \includegraphics[width=0.48\textwidth]{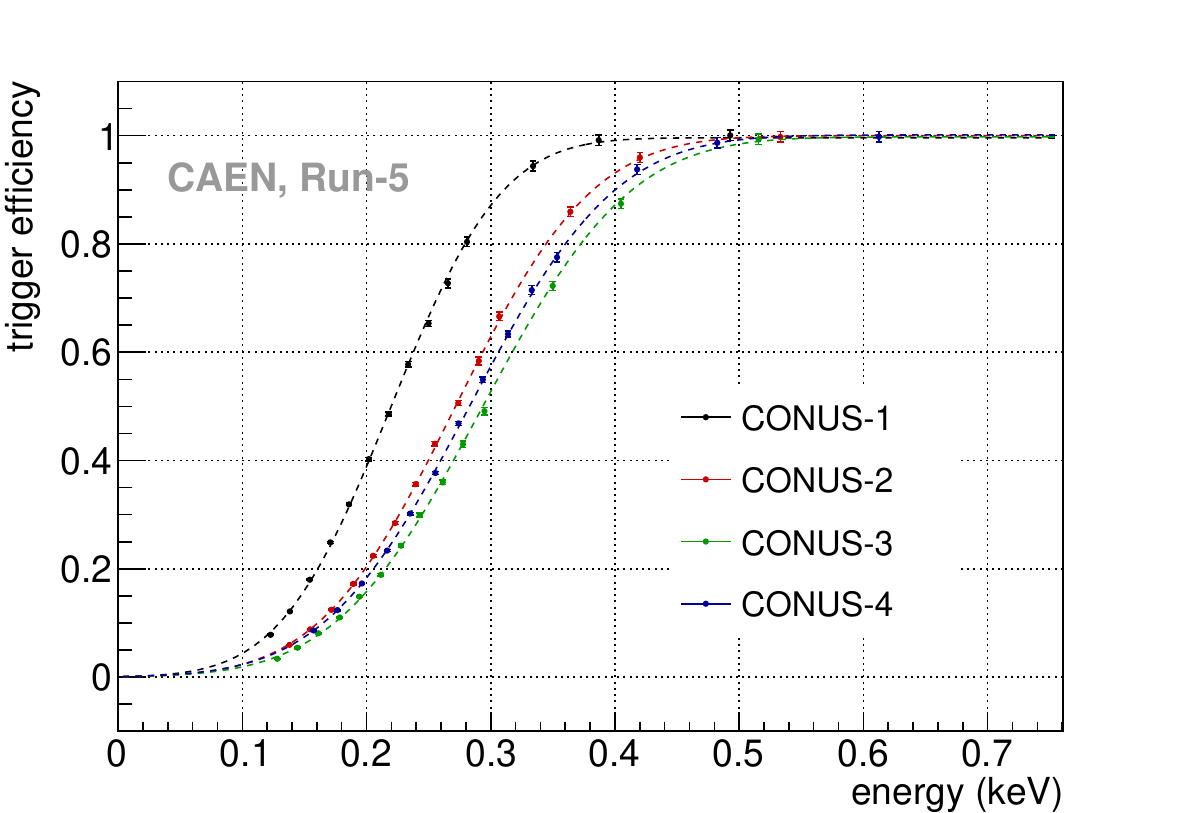}
    \caption{Measured trigger efficiencies for the four \textsc{Conus} detectors (points) with the CAEN DAQ during \textsc{Run-5} and associated best-fit description (dashed lines) using the analytical description of Eq. (\ref{eq:trigger_efficiency}).}
    \label{fig:trigger_efficiency_curves}
\end{figure}

A data acquisition system (DAQ) system based on commercially available CAEN digitizers offering pulse processing implemented at the FPGA level was installed for the data collection of \textsc{Conus} \textsc{Run-5}, lasting from May 21, 2021 until December 5, 2022 along with the pre-existing Lynx DAQ system.
The CAEN DAQ is operated via the CoMPASS software from CAEN and data are further preprocessed and analysed offline.
The HPGe signals from the four \textsc{Conus} detectors were sampled at a rate of 100\,MHz rate by a CAEN V1782 multi-channel analyzer. Inputs with a 10\,$\mu$s AC coupling were used in order to get rid of the typical TRP step-like rising baseline level.
A combination of a slow and a fast triangular discriminator was used for the trigger. This gave better results in terms of detection efficiency of small signals compared to the standard RC-CR$^2$ algorithm. The fast and slow triangular discriminators have shaping times of 0.8\,$\mu$s and 2.4\,$\mu$s, respectively. The discriminator thresholds were set as low as possible to optimize the detection efficiency of low signals without exceeding 1\,\% of dead-time. This results in noise trigger rates of about 500\,Hz. The trigger efficiency was determined by injecting artificial signals produced by a pulse generator with the same rise time as of the physical signals. Fine-grained scans allowed to precisely measure the detector response as a function of the energy. They are shown for the four detectors in Fig.\,\ref{fig:trigger_efficiency_curves}. The \textsc{Conus-1} detector exhibits a better trigger efficiency due to its slightly faster signal rise time combined with a lower noise level compared to the other three detectors. A suitable description of the experimental trigger efficiency curves was obtained using the fit function:
\begin{equation}\label{eq:trigger_efficiency}
    \varepsilon_{trig} = 0.5\cdot\left(1+\text{erf}\left(\frac{E_{ee}-t_1}{t_2}\right)\right)\, ,
\end{equation}
with the error function erf and $t_1$ being in $(225-295)\,\text{eV}_{\text{ee}}$ and $t_2$ in $(105-138)\,\text{eV}_{\text{ee}}$, respectively, depending on the detector. Note that the measured trigger efficiencies with this DAQ are slightly worse compared to the ones obtained with the pre-existing Lynx DAQ system \cite{Bonet2021d} but this configuration was favored since it was found to offer a better compromise between accepted noise events and signal efficiency, resulting in a better signal-over-background ratio in the region of interest for the \CEvNS~search.

The energy of each event was reconstructed by a trapezoidal shaping filter with a shaping time of 12\,$\mu$s and a trapezoidal flat top of $(3-6)\,\mu\text{s}$ depending on the detector, optimized in terms of energy resolution. Signals separated by less than the trapezoidal shaping time are identified as pile-up events and are rejected from the analysis chain. Because of the very low count rates of the \textsc{Conus} detectors, the typical encountered pile-up rates during data collection were very low -- about 2\,Hz. The logic signals from the $\mu$-veto system as well as the TRP reset signals from the Ge detectors were acquired with a CAEN V1725 module. The two modules shared a common synchronization clock. All triggering events were saved and the time cuts ($\mu$ and TRP) were applied offline in the analysis chain. In addition, the raw waveforms in 20\,$\mu$s windows are saved for all events triggering the Ge detectors and are used for the offline pulse shape studies presented in this article.

\section{Methods} \label{section:methods}

\subsection{Rise time fit}\label{subsection:fit}
 
\begin{figure*}
		\centering
		\begin{subfigure}[c]{0.45\textwidth}
			\includegraphics[width=\textwidth]{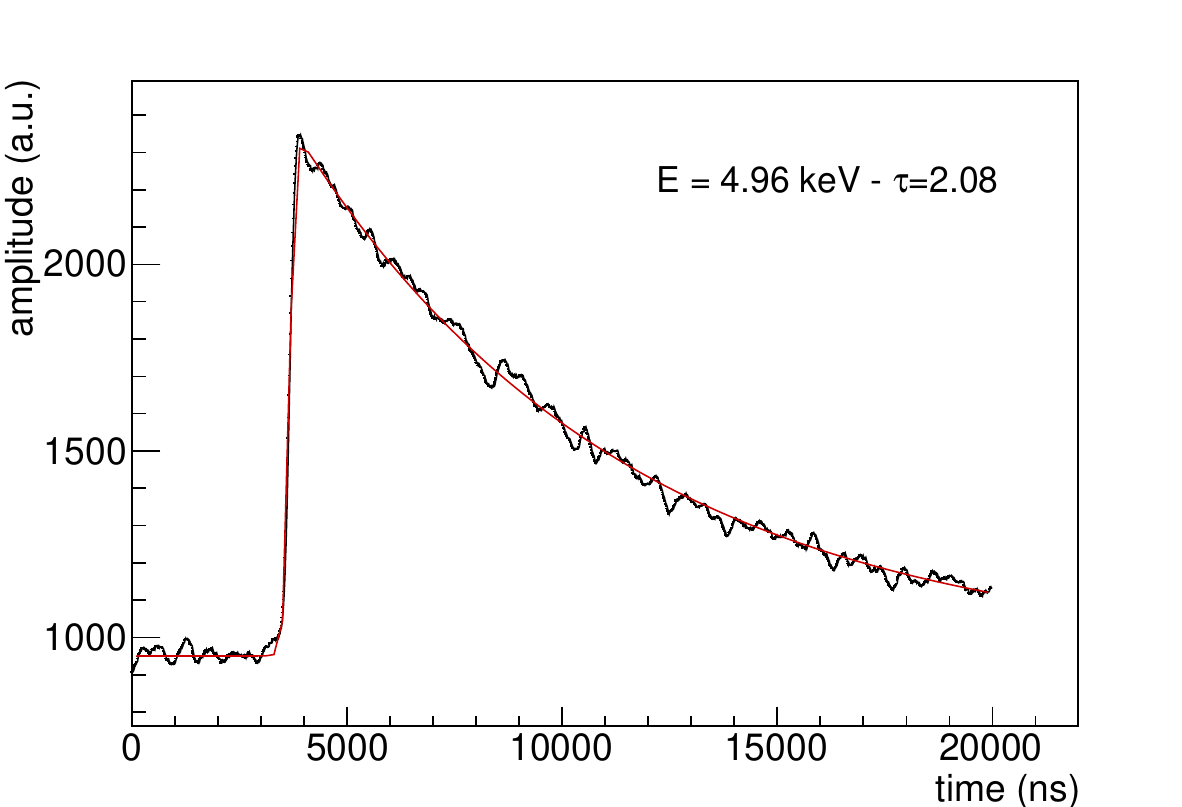}
			\caption{}\label{fig:c1_np_5000eV}
		\end{subfigure}
		\begin{subfigure}[c]{0.45\textwidth}
			\includegraphics[width=\textwidth]{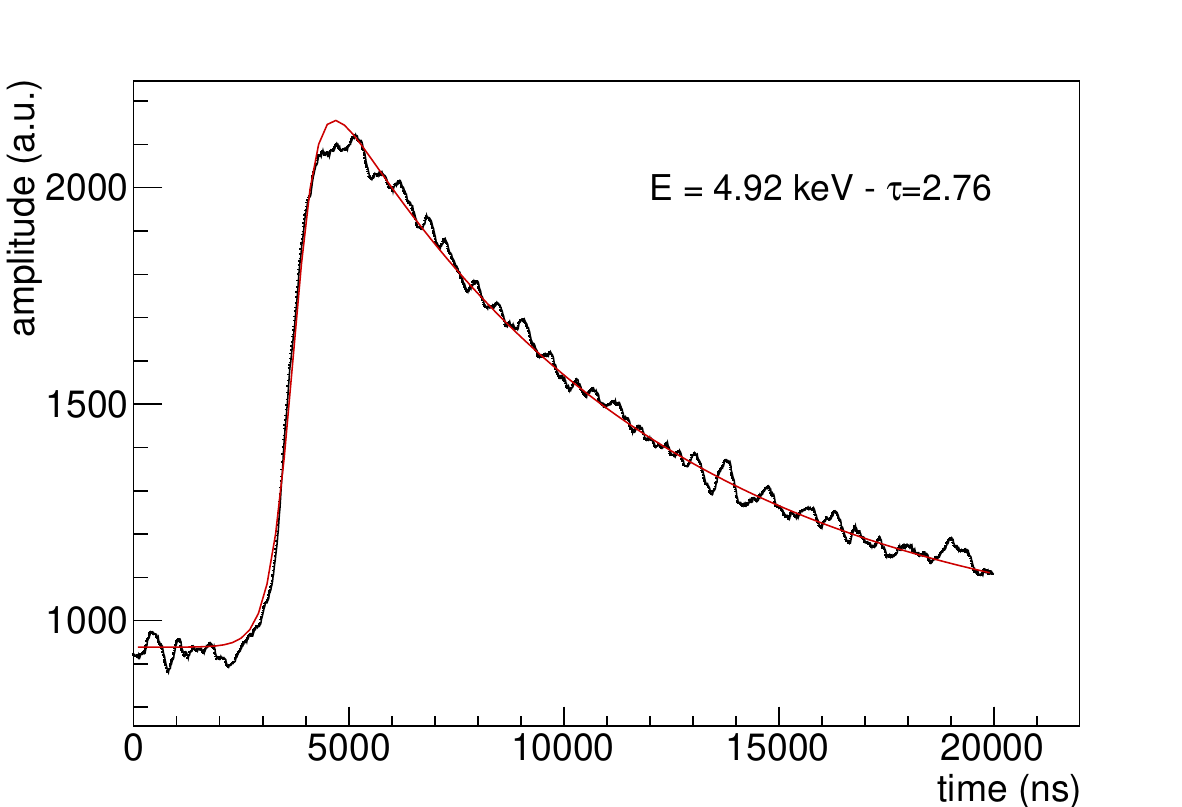}
			\caption{}\label{fig:c1_sp_5000eV}
		\end{subfigure}
		\begin{subfigure}[c]{0.45\textwidth}
			\includegraphics[width=\textwidth]{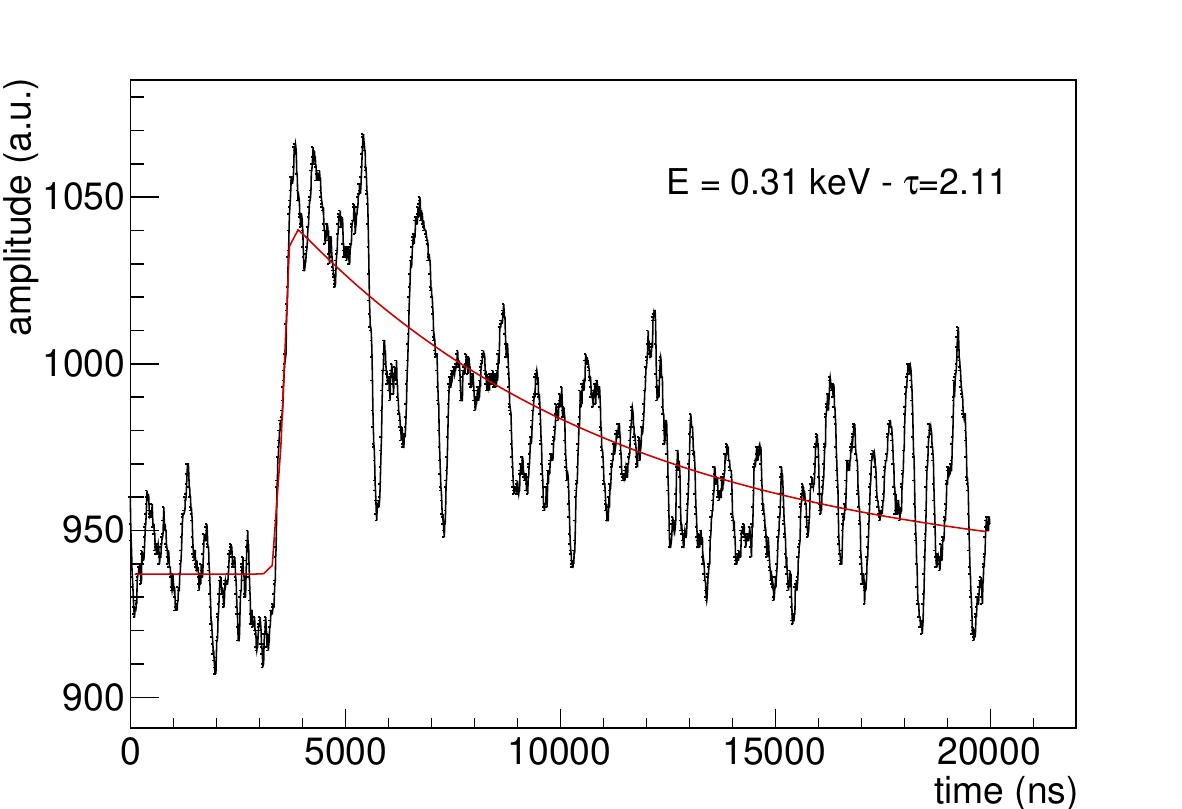}
			\caption{}\label{fig:c1_np_300eV}
		\end{subfigure}
		\begin{subfigure}[c]{0.45\textwidth}
			\includegraphics[width=\textwidth]{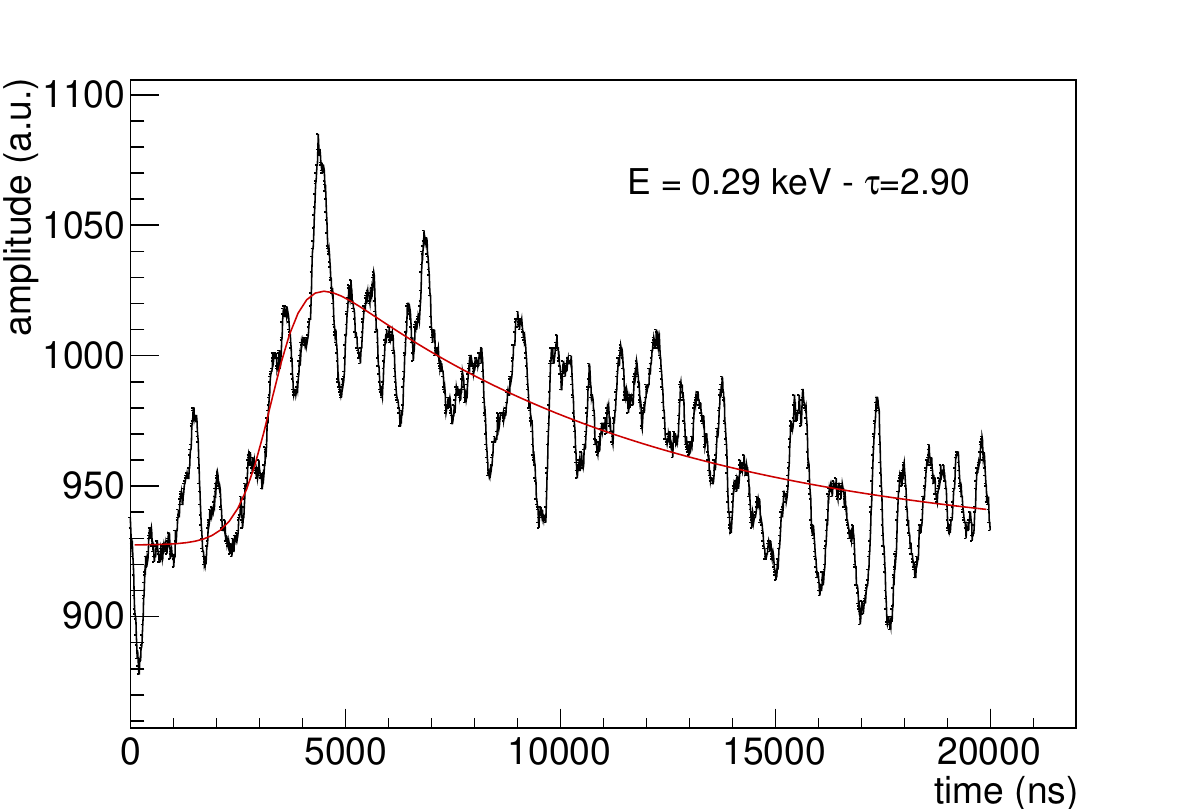}
			\caption{}\label{fig:c1_sp_300eV}
		\end{subfigure}
		
		\caption{Fit examples of pulses at high (top) and low (bottom) energies for a FP (left) and a SP (right) of the \textsc{Conus-1} detector.\label{fig:pulses_fits}}
\end{figure*}

The physical information allowing to discriminate bulk from surface events is contained in the $\mathcal{O}$(100\,ns) rise time of the signals and was already exploited for PPC HPGe by CoGENT \cite{Aalseth2013} and TEXONO \cite{Yang2018b}. This rise time can be extracted on an event-by-event basis by fitting the waveforms with the following analytical description:
\noindent
\begin{equation}
    f(t) = A_{0}\left[\text{tanh}(\frac{t-t_0}{\tau})+1\right]\text{exp}(-\tau_c(t-t_0)) + P_0\, ,
\end{equation}\label{eq:fit}
where $A_{0}$ is the amplitude of the signal, $t_{0}$ the timing offset, $P_{0}$ the baseline noise level and $\tau$ the parameter of interest related to the rise time of the event. The exponential decay accounts for the AC coupling of the DAQ with a characteristic time $\tau_{C}\,\sim\,(7.5-8)\mu$s. This parameter, being a constant of the system, was fixed in the fit in order to help the fit converging at low energy. Exemplary fits for FPs induced by bulk events and SPs induced by surface events are shown in Fig.\,\ref{fig:pulses_fits}. For all four \textsc{Conus} detectors the rise time $\tau$ equals $(120-180)\,\text{ns}$ for the FP (equivalent to t$_{10-90\,\%}$ of $(275-325)\,\text{ns}$). This rise time results both from the charge collection time in the diode and from the time response of the preamplifier. The difference between the detectors is explained by different settings of the preamplifiers. Depending on the nature of the energy deposition, SPs are observed with rise times up to $\sim$\,1\,$\mu$s, justifying for visibility reasons the logarithmic scale used for most of the rise time distributions shown in this article.

Complementary methods (two sub-types of the A/E method \cite{Budjas2009} and an integral-ratio method) were tested but were shown to be less efficient below a few keV$_{ee}$ compared to the $\tau$ fit method. Alternative parametrisations were also tested for the pulse description and the use of the hyperbolic tangent in Eq.\,(\ref{eq:fit}) was found to be the best compromise between computation time and discrimination power \cite{Henrichs2021}.

\subsection{Physical event samples}\label{subsection:physics_samples}

For the energy calibration and the study of the shape of the pulses, advantage is taken of the complementarity of various event samples. 

First, calibration data from a $^{228}$Th source emitting $\gamma$-rays up to 2.6\,MeV are available. These measurements were repeated every 2--4 weeks during the whole data collection period allowing to monitor the stability of the rise time over time. The $\sim$15\,kBq source is inserted inside the \textsc{Conus} shield \cite{Bonet2021} at a 20\,cm distance from the diodes allowing for reasonable statistics in a few hours and provides a good mixture of single-site events (SPs, FPs and VFPs), MS and pile-up events.

Second, the \textsc{Conus} background data were considered. Before the application of any veto, the experiment is dominated by muon-induced backgrounds. Hence, data recorded in reactor-OFF before the application of the $\mu$-veto cut provide a high-statistics sample of $\mu$-induced events which can be split in an electromagnetic and a neutron-induced component \cite{Bonet2021_bkg}. In the region of interest below 10\,\keVee, the first component is sub-dominant and is expected to contribute more to SP events, while the second dominant component consists mainly of neutron-induced recoils inducing FP bulk events given the large mean free-path of the neutrons.
The application of the $\mu$-veto cut suppresses $\sim$\,97\,\% of this background and the remaining $\mu$-induced components are found to be of the same order of magnitude as for the detector-dependent intrinsic contamination of $^{210}$Pb, present in the soldering in the vicinity the diodes. Both type of backgrounds can result in SP and FP energy depositions. However and according to the MC background model, a significant amount of the $^{210}$Pb decays are expected to occur close to and even on the surface of the diode close to the point contact, which is covered by the passivation layer (cf. Fig.\,\ref{fig:diode}). Therefore, these decays will produce almost no SPs: as there is no transition layer, they will induce energy deposition within the bulk volume in the close vicinity of the p+ contact without experiencing partial charge collection.

Finally, neutron activation in the germanium itself is responsible of the appearance of characteristic lines in the background spectrum. The K-shell line at 10.37\,keV and L-shell line at 1.30\,keV originating from atomic de-excitation after electron capture in $^{68,71}$Ge are in particular well visible in the \textsc{Conus} background, with a counting rate of about 16\,counts/d/kg (2\,counts/d/kg) for the K-shell (L-shell) during \textsc{Run-5}. \textit{In situ} activation of $^{71}$Ge is the main contribution due to the long storage of the detectors inside the shield\cite{Bonet2022}. The corresponding events are particularly interesting since they are expected to be homogeneously distributed within the full Ge volume and to contribute mainly to FP events in contrast to the complex vertices distribution of the background. A clean sample of these events can be used to reproduce artificially FP events, see Sec. \ref{subsection:artificial_bulk_generation}. 

\begin{figure}
    \centering
    \includegraphics[width=0.48\textwidth]{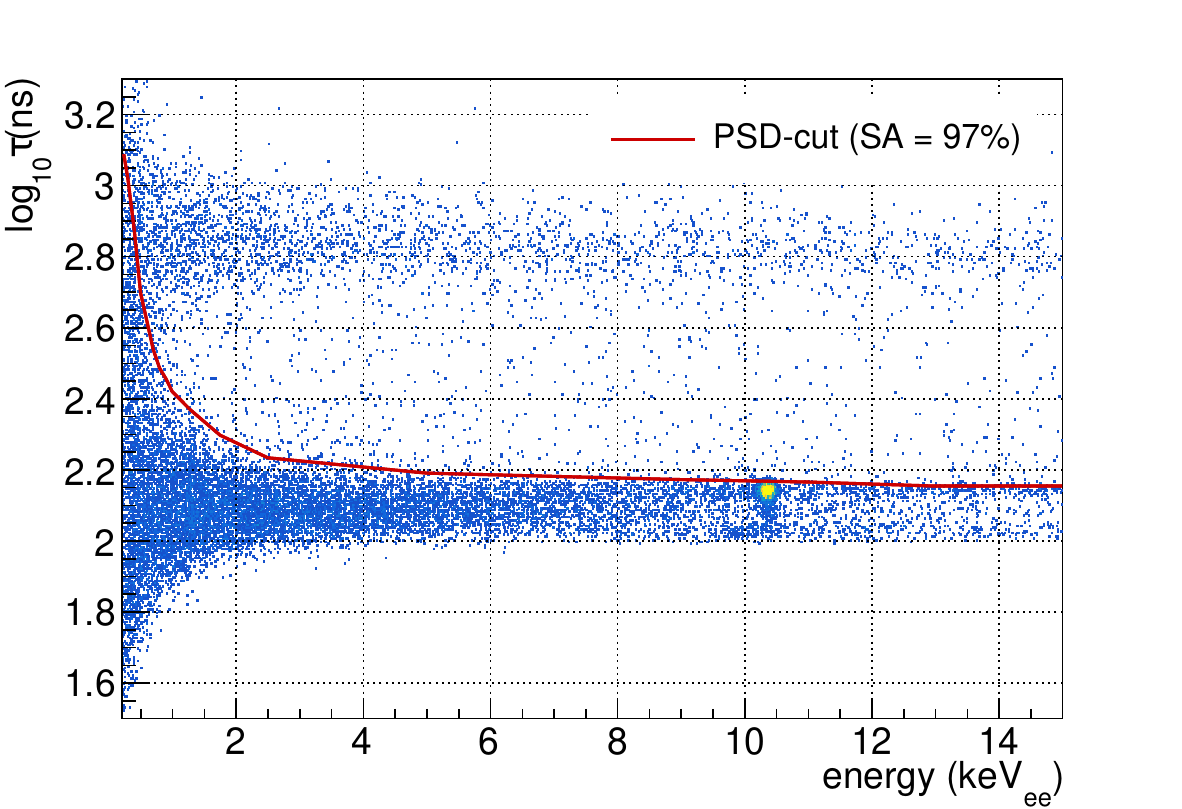}
    \caption{Rise time of background events in dependence of energy after application of the $\mu$-veto for the \textsc{Conus-1} detector. Bulk events are characterised by a fast rise time and populate the lower band. The 10.37\,keV Ge activation line is clearly visible within the bulk event population. Surface events produce slow signals which are found in the upper band. The red line shows exemplary a cut with a 97\,\% FP signal acceptance (SA).}
    \label{fig:bkg_2D}
\end{figure}

The rise time of background events recorded during reactor-OFF after application of the $\mu$-veto is shown in Fig.\,\ref{fig:bkg_2D} as a function of energy. The fast and slow populations (around log$_{10}\tau\sim2.1$ and log$_{10}\tau\sim2.9$, respectively) are clearly separated for the highest energies and start to mix in the sub-\keVee~regime due to the growing influence of electronic noise. The gap between the two main bands is populated by MS or pile-up events which are too close to be discarded by the pile-up rejection of the DAQ and whose pulse fit leads to artificially slow rise times.
The lower band gathers both the FPs (log$_{10}\tau\gtrsim2.1$) resulting from interactions inside the bulk volume and VFPs (log$_{10}\tau\lesssim2.1$) for interactions close to the p+ contact as illustrated in Fig.\,\ref{fig:three_pulses}. The 10.37\keVee~K-shell atomic de-excitation line following Ge EC decays is particularly visible within the FP part of the band since for these homogeneously distributed events, only the small fraction of the decays close to the p+ contact are expected to lead to VFPs.

\begin{figure*}
		\centering
		\begin{subfigure}[c]{0.48\textwidth}
			\includegraphics[width=\textwidth]{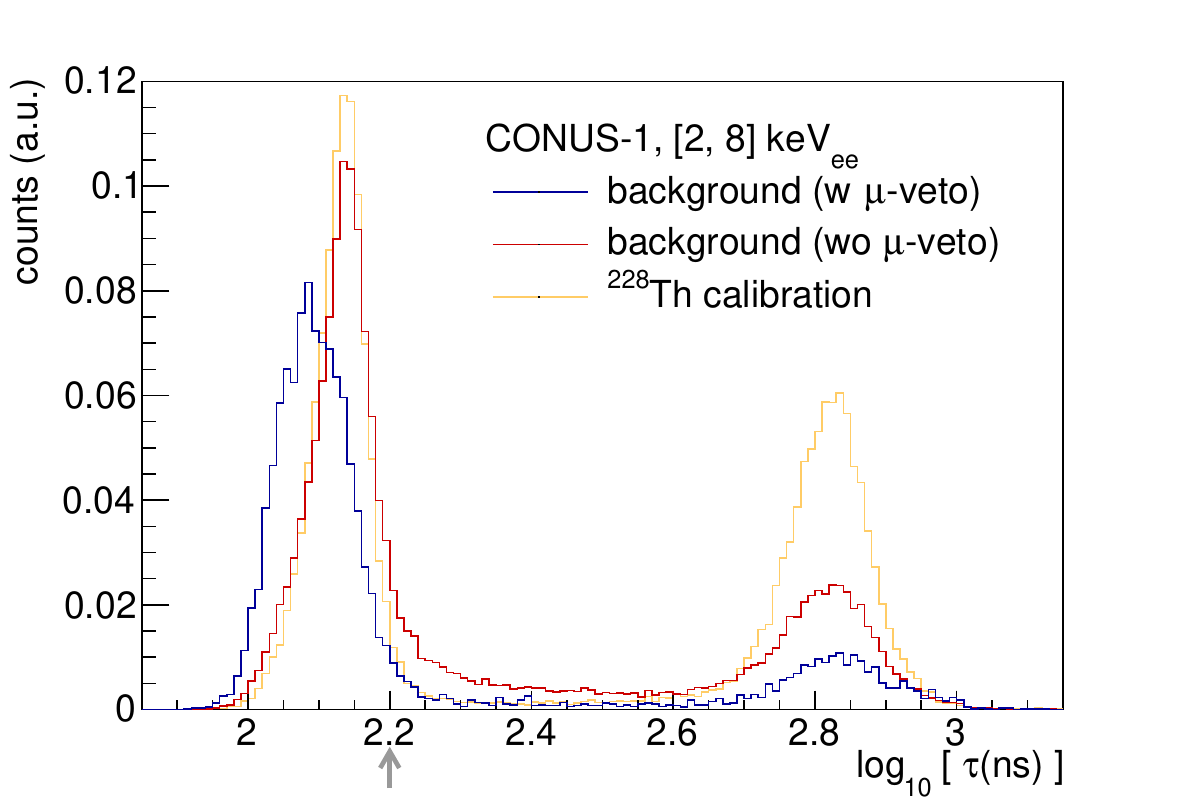}
			\caption{}\label{fig:psd_populations_c1_highE}
		\end{subfigure}
		    \begin{subfigure}[c]{0.48\textwidth}
			\includegraphics[width=\textwidth]{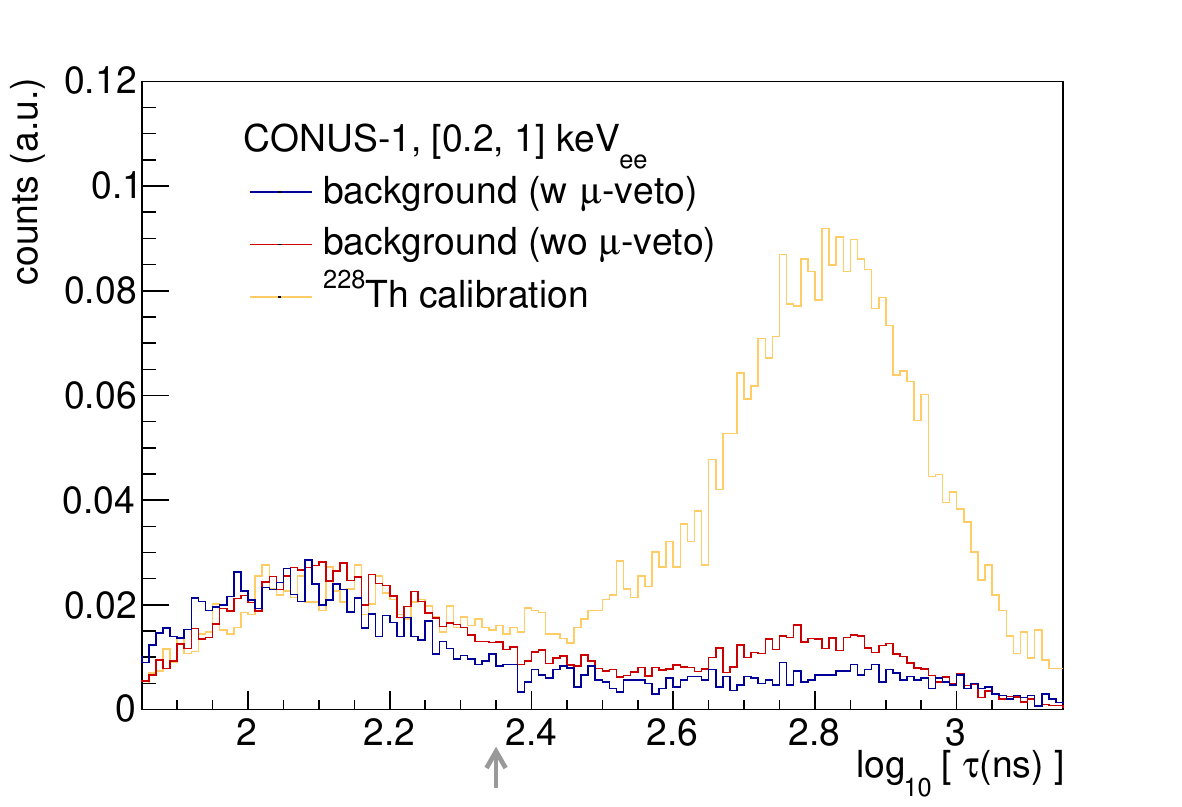}
			\caption{}\label{fig:psd_populations_c1_lowE}
		\end{subfigure}
		\begin{subfigure}[c]{0.48\textwidth}
			\includegraphics[width=\textwidth]{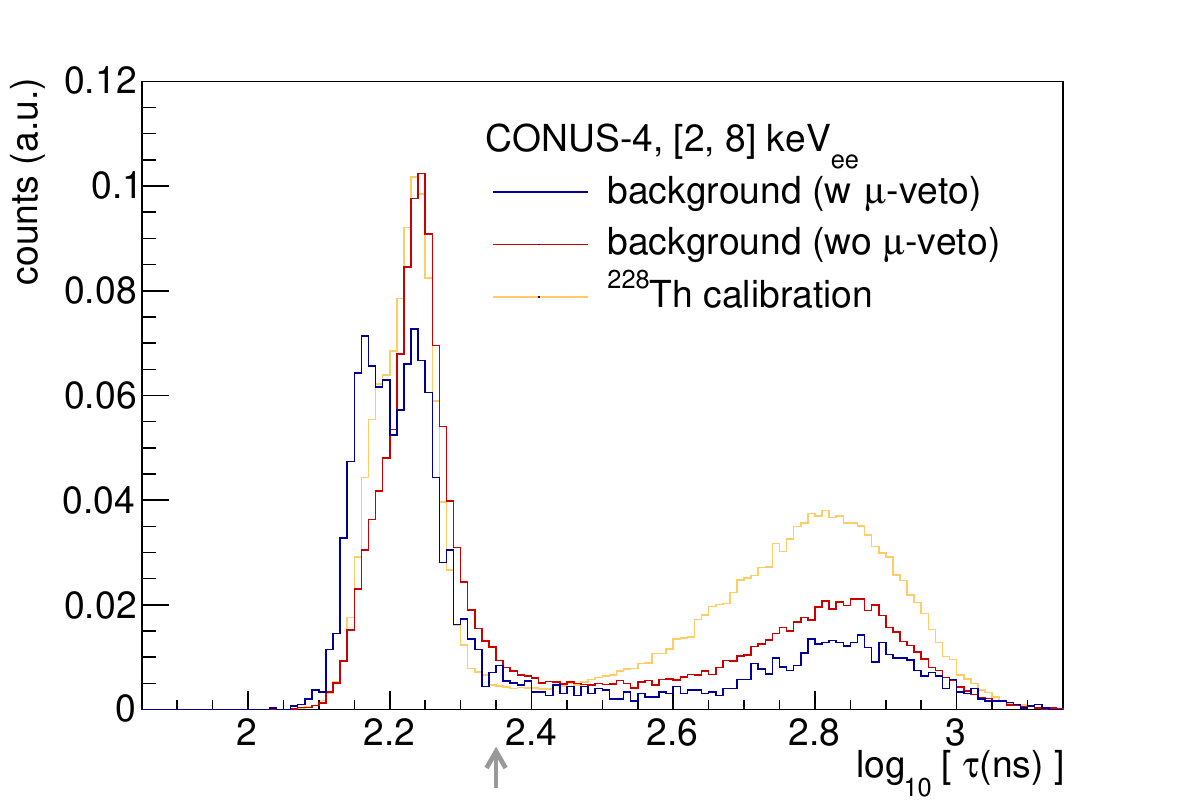}
			\caption{}\label{fig:psd_populations_c4_highE}
		\end{subfigure}
		\begin{subfigure}[c]{0.48\textwidth}
			\includegraphics[width=\textwidth]{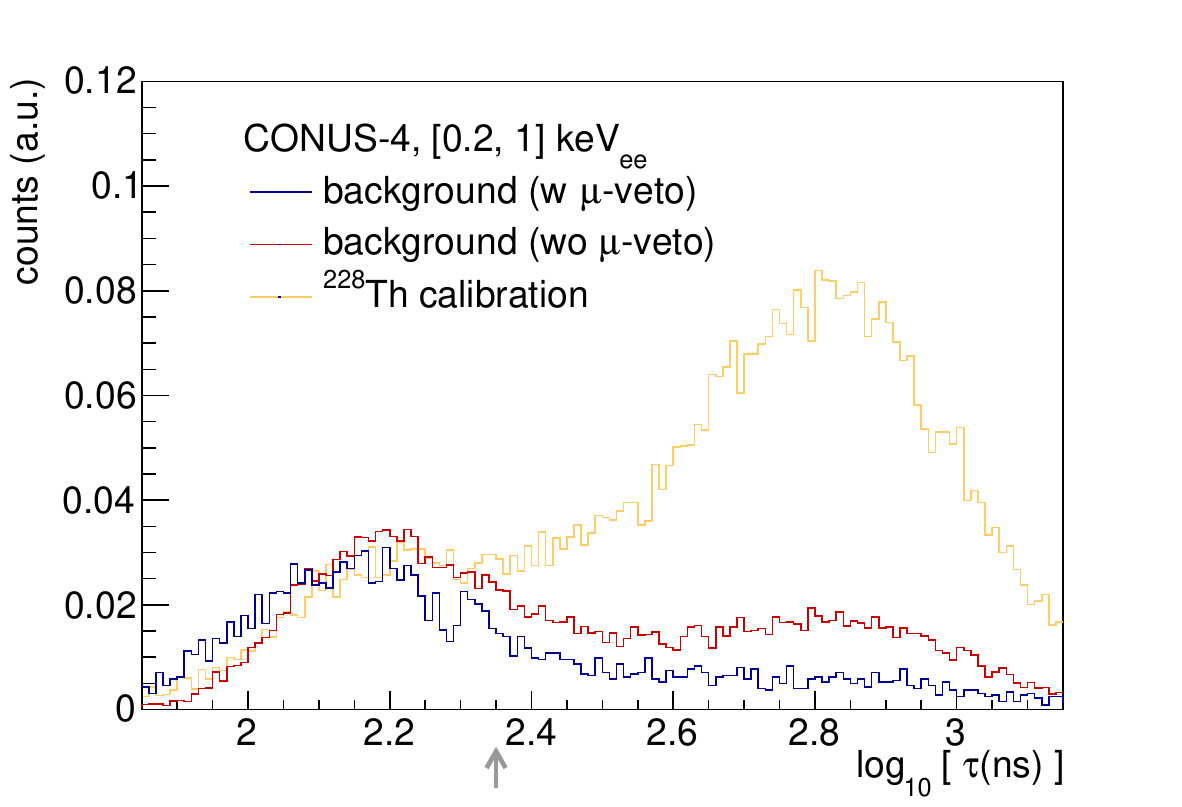}
			\caption{}\label{fig:psd_populations_c4_lowE}
		\end{subfigure}
		\caption{Rise time distributions for different event populations and for two detectors in two different energy ranges: (a): \textsc{Conus-1} for energies in (2,\,8)\keVee. (b): \textsc{Conus-1} for energies in (0.5,\,1)\keVee. (c): \textsc{Conus-4} for energies in (2,\,8)\keVee. (d): \textsc{Conus-4} for energies in (0.5,\,1)\keVee. Background data collected in reactor-OFF before (after) application of the $\mu$-veto cut are shown in red (blue), and calibration data from a $^{228}$Th radioactive source irradiation are given in orange. The upper range used for normalisation is indicated by the gray arrow.\label{fig:psd_populations} }
\end{figure*}

To illustrate the complementary in terms of pulse shape between the aforementioned physical samples, Fig.\,\ref{fig:psd_populations} presents their rise time distributions at mid (left pannel) and low energy (right panel) for two exemplary \textsc{Conus} detectors (similar distributions as for \textsc{Conus-4} are obtained for the two others). From this it can be seen that the discrimination power between the two main populations (SPs and FPs) is detector dependent: \textsc{Conus-1} exhibits a better separation than the other \textsc{Conus} detectors thanks to its slightly faster response.
Information on the underlying background components can be gained by comparing the FP distributions at high energy before and after the application of the $\mu$-veto, in red and blue, respectively. It appears that the remaining background with applied $\mu$-veto contains a VFP component -- negligible without $\mu$-veto -- dominating the distribution for \textsc{Conus-1} and of equal weight as the FPs for \textsc{Conus-4}. This is consistent with the findings about the background decomposition detailed in \cite{Bonet2021_bkg}, showing that the background of \textsc{Conus-1} is dominated by the intrinsic $^{210}$Pb contamination (responsible for the VFP component since the decays occur mainly in the p+ contact region according to the MC model) whereas the tendency is inverted for the other detectors. 

At low energy however, the two populations mix due to the growing influence of electronic noise. The large difference in terms of fraction of SPs between the $^{228}$Th calibration data and the background data confirms the high-dependency of vertex distributions.

\subsection{Artificial generation of bulk signals}\label{subsection:artificial_bulk_generation}
 
In order to determine the cut acceptance for neutrino events, the rise time distribution of homogeneously distributed bulk events has to be known. As mentioned above, no calibration source allows to have a pure homogeneous bulk event sample over the whole energy range. To overcome this obstacle, artificial signals from an electronic pulser mimicking bulk events were generated and injected through the whole DAQ chain and treated in the same way as measured physical signals.

\begin{figure}
    \centering
    \includegraphics[width=0.48\textwidth]{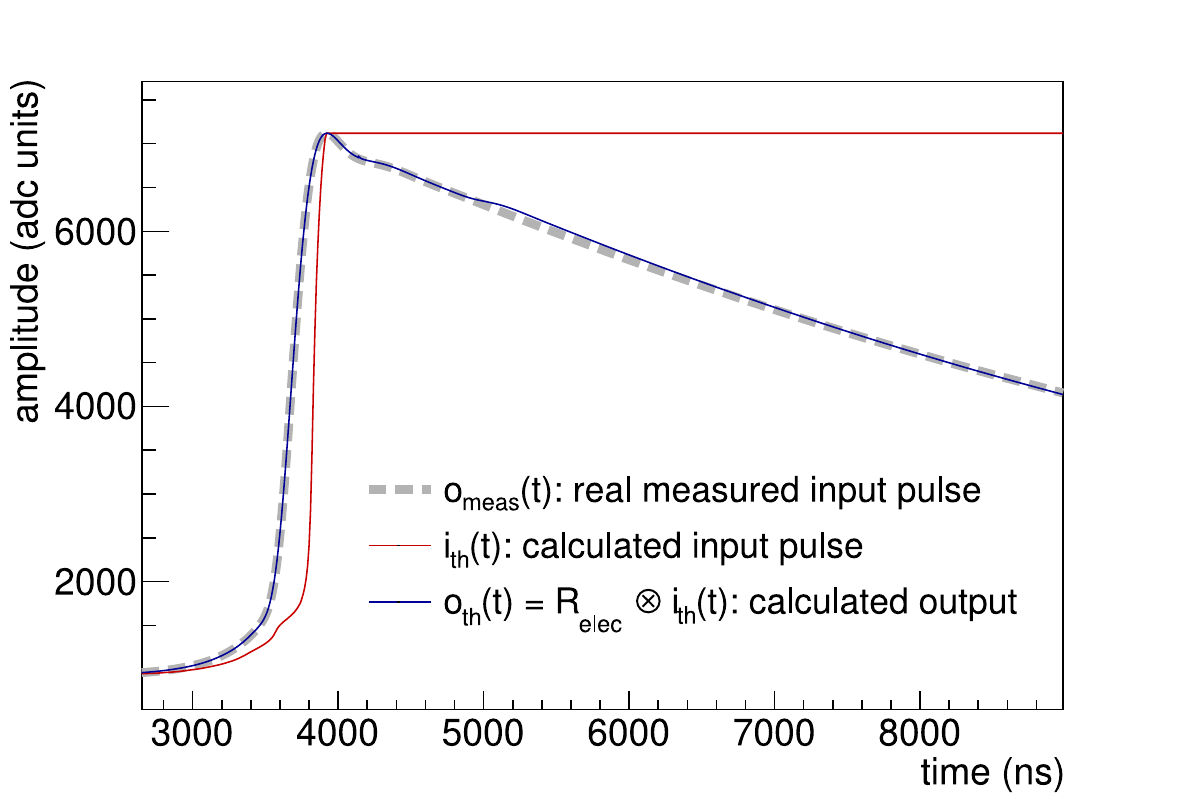}
    \caption{Illustration of the iterative method: starting from a high-energy measured real pulse ($o_{meas}(t)$, gray dotted line), the unknown input pulse corresponding to bulk events ($i_{th}(t)$, red line) is determined iteratively such that its convolution with the detector response ($o_{meas}(t)$, blue line) matches the real pulse.}
    \label{fig:iterative_method}
\end{figure}

The correct shape of these signals can be determined either via a validated electric field simulation of the diode or via a data-driven method consisting in deconvolving the measured bulk signals at higher energy from the DAQ chain response, the latter was chosen in this work. The deconvolution was performed in a semi-empirical way as follows: the impulse response function of the DAQ chain was first measured independently for all detectors by injecting a fast rectangular signal in the test input of the preamplifier. The input signal mimicking a bulk signal was then determined iteratively as the signal whose convolution with the impulse response function matches a real measured physical output pulse. In practise, a measured signal was used as initial pulse and was distorted in an iterative way to achieve a match to the output signal. About ten iterations are sufficient to obtain a very good agreement ($\lesssim$\,1\,ns difference in terms of rise time) between the convolution of the calculated input pulse and the real pulses \cite{Stauber2022}.
This procedure is illustrated in Fig.\,\ref{fig:iterative_method}. The obtained red signal is then used as description of a pure bulk signal and injected in the DAQ chain of the detectors for the determination of the rise time distribution. The energy dependence of the distribution was studied by a simple scaling of this input signal. The hypothesis of a constant shape independent of the energy was verified using calibration data between 1 and 10\keVee.

\begin{figure}
    \centering
    \includegraphics[width=0.48\textwidth]{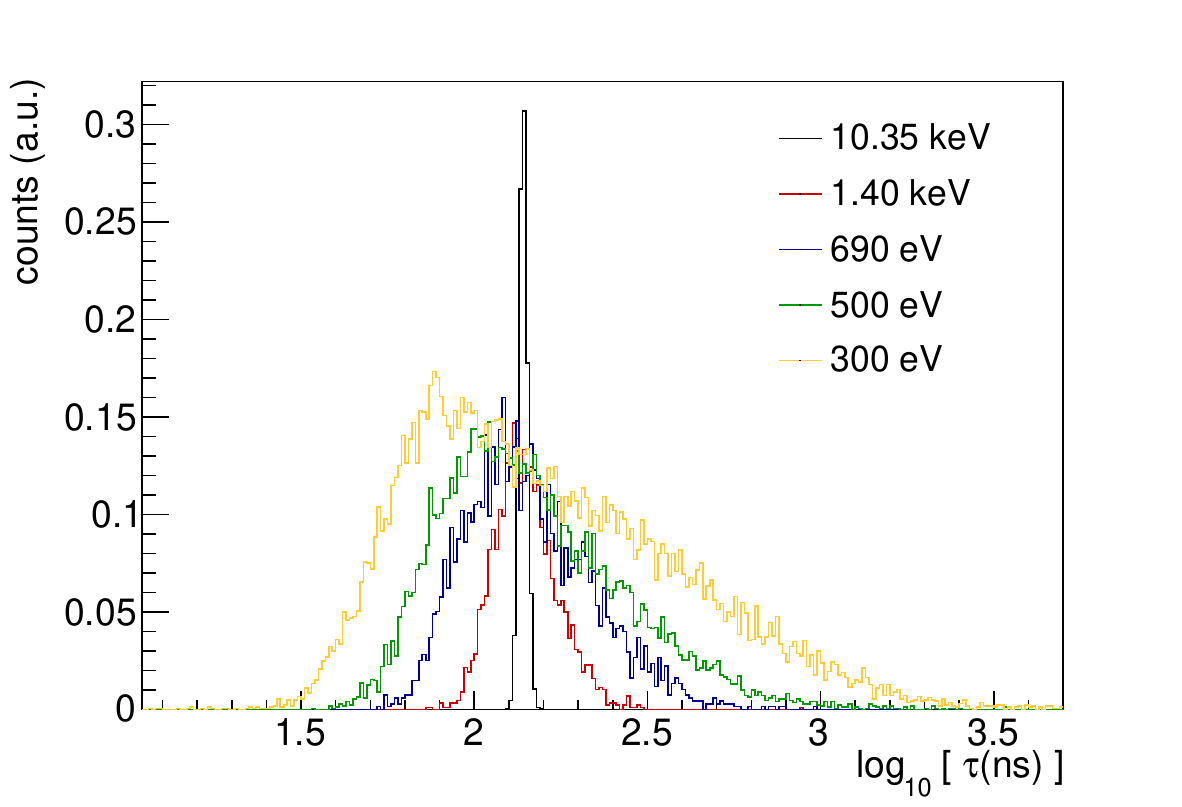}
    \caption{Rise time distributions from artificially generated bulk events for different energies, from 300\,eV$_{ee}$ to 10.35\,\keVee. The electronic noise is responsible for the broadening of the distributions and for the deviation from a Gaussian below 1\,\keVee.}
    \label{fig:psd_pulser_energies}
\end{figure}

Fig.\,\ref{fig:psd_pulser_energies} illustrates the evolution of the measured rise time distributions as a function of the energy obtained from a scan over the amplitude of the theoretical input signal mimicking bulk events for \textsc{Conus-1}. As the energy decreases, the relative importance of noise increases and explains the larger smearing of the distributions. Note that for energies $\lesssim$\,1\keVee, the distributions also deviate from a Gaussian shape and exhibit an asymmetric behavior. From these distributions, the signal acceptance (SA) of FPs can be estimated for any rise time cut value.
At very low energy ($\lesssim$\,0.4\keVee), a small fraction of the rise time fits do not converge and the corresponding events are discarded from the analysis chain, leading to an additional reduction in the SA. It amounts to about 1\,\% at 300\,eV$_{\text{ee}}$ and 3\,\% at 250\,eV$_{\text{ee}}$.

\section{Results and discussion} \label{section:results}

\subsection{Validation and systematic uncertainties}\label{subsection:validation}

The four physics samples described in Sec.\,\ref{subsection:physics_samples} can be used to validate qualitatively and quantitatively the correct classification of the pulses via their rise time.

First, the rise time distribution of the $^{228}$Th calibration data were compared to a MC simulation (see \cite{Bonet2022} for a more detailed description of the $^{228}$Th MC simulations) in order to validate the ratio of bulk events over surface events. This ratio depends on the energy of the $\gamma$-rays, the width of the detector-dependent transition layers and the geometry of the full experimental setup. From a \textsc{Geant4} MC simulation of such a calibration measurement, the fraction of interactions happening in the transition layers can be determined and compared to the amount of pulses identified as slow in the data. The fraction of SPs is about 20\,\% above 15\,keV and increases to 30\,\% (45\,\%) towards low energy, in the $(6-12)\,\text{keV}$ and $(2-6)\,\text{keV}$ ranges, respectively, due to the larger attenuation. An agreement at the 5\,\% level is achieved \cite{Henrichs2021}. This demonstrates on one hand the accurate description of the detector transition layer and on the other hand the consistency of the surface event identification over the energy, showing no additional unexpected component within this population. A percent-level comparison was achieved by coupling the vertex distributions from the MC with a full pulse shape simulation, allowing to take also into account MS events corresponding to multiple energy depositions within one event, that lead to an artificially larger fitted rise time if they occur within a too small time interval.

\begin{figure}
    \centering
    \includegraphics[width=0.48\textwidth]{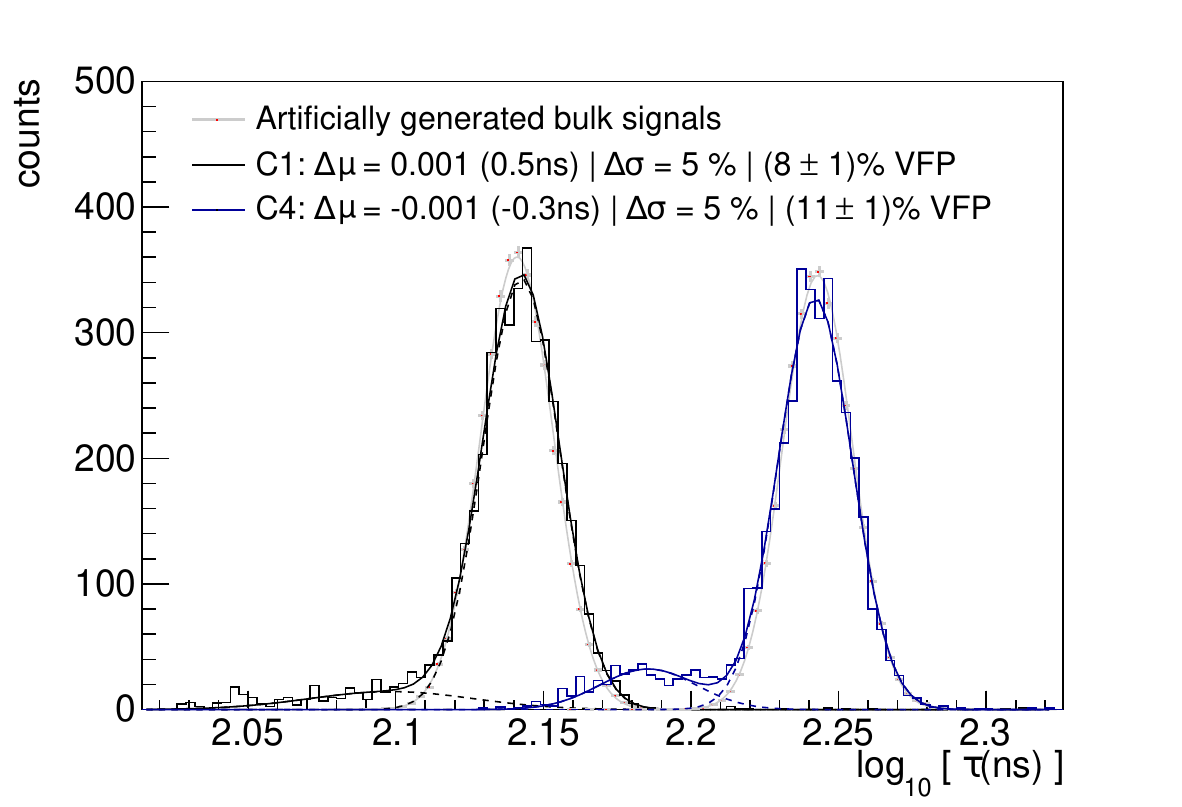}
    \caption{Rise time distribution of events in the 10.37\,keV line after background subtraction for \textsc{Conus-1} (black) and \textsc{Conus-4} (blue).
    The sub-component with faster rise time corresponds to the VFPs, which correspond to energy depositions in the bulk region close to the p+ contact. The rise time distribution of the artificially generated fast bulk signals is shown in gray and is in agreement with the measured distributions. The fraction of VFPs is extracted via a fit of the data with two Gaussian distributions modeling the two components.}
    \label{fig:validation_10keV_ref}
\end{figure}

Second, as a validation of the procedure of the artificial generation of bulk events described in Sec.\,\ref{subsection:artificial_bulk_generation}, the measured rise time distributions resulting from the injected theoretical bulk signal were compared with the sample of real physical events from the 10.37\,\keVee~line from Ge activation. As mentioned above, these events are expected to be homogeneously distributed in the diode volume and should in consequence be predominantly made of bulk events. Their distribution is obtained statistically after subtraction of the background under the line and are shown exemplary for \textsc{Conus-1} and \textsc{Conus-4} in Fig.\,\ref{fig:validation_10keV_ref}. The remaining sub-component with faster rise time values corresponds to VFPs originating from bulk events close to the p+ contact. The fast component of the distributions is compared to the rise time distribution of the artificial bulk signals, also shown in gray. By construction, an agreement in terms of mean rise time is achieved at the ns level. Most importantly, the width of the distributions is found to be in agreement at the $(5-10)\,\%$ level, indicating that the influence of noise is mostly responsible for the smearing of the distribution and that no additional process contributes significantly to the variability in the shape of the physical bulk pulses. The small remaining differences -- both in terms of mean value and width of the distributions as reported in Tab.\,\ref{tab:data_conus_detectors} -- are taken as systematic uncertainties, their impact is discussed below.

\begin{table*}
    \centering
    \begin{tabular}{c|c c | c | c c }
         Detector & $\Delta\mu$ (log$_{10}$) & $\Delta\mu$ (ns) & $\Delta\sigma$ (\%) & VFP fraction (\%) & r$_{p+}$ (mm)\\
         \hline
         \textsc{Conus-1} & 0.001 & 0.5 & 5 &  8 $\pm$ 1 & 19.5 $\pm$ 0.8 \\
         \textsc{Conus-2} & 0.002 & 0.8 & 4 & 13 $\pm$ 1 & 22.7 $\pm$ 0.6 \\ 
         \textsc{Conus-3} & 0.004 & 1.7 &10 & 12 $\pm$ 2 & 22.1 $\pm$ 1.2 \\ 
         \textsc{Conus-4} &-0.001 &-0.3 & 5 & 11 $\pm$ 1 & 21.7 $\pm$ 0.2 \\ 
    \end{tabular}
    \caption{Observed deviation of mean ($\Delta\mu$) and width ($\Delta\sigma$) of the rise time distributions of FPs events originating from the artificially pulser-generated bulk signals and the homogeneous 10.37\,keV data event sample. The fraction of VFPs can also be extracted from these two data distributions. They can be directly translated into an effective half-sphere volume around the p+ contact of radius r$_{p+}$ in which the charge collection happens \textit{very fast.} }
    \label{tab:data_conus_detectors}
\end{table*}

\begin{figure}
    \centering
    \includegraphics[width=0.48\textwidth]{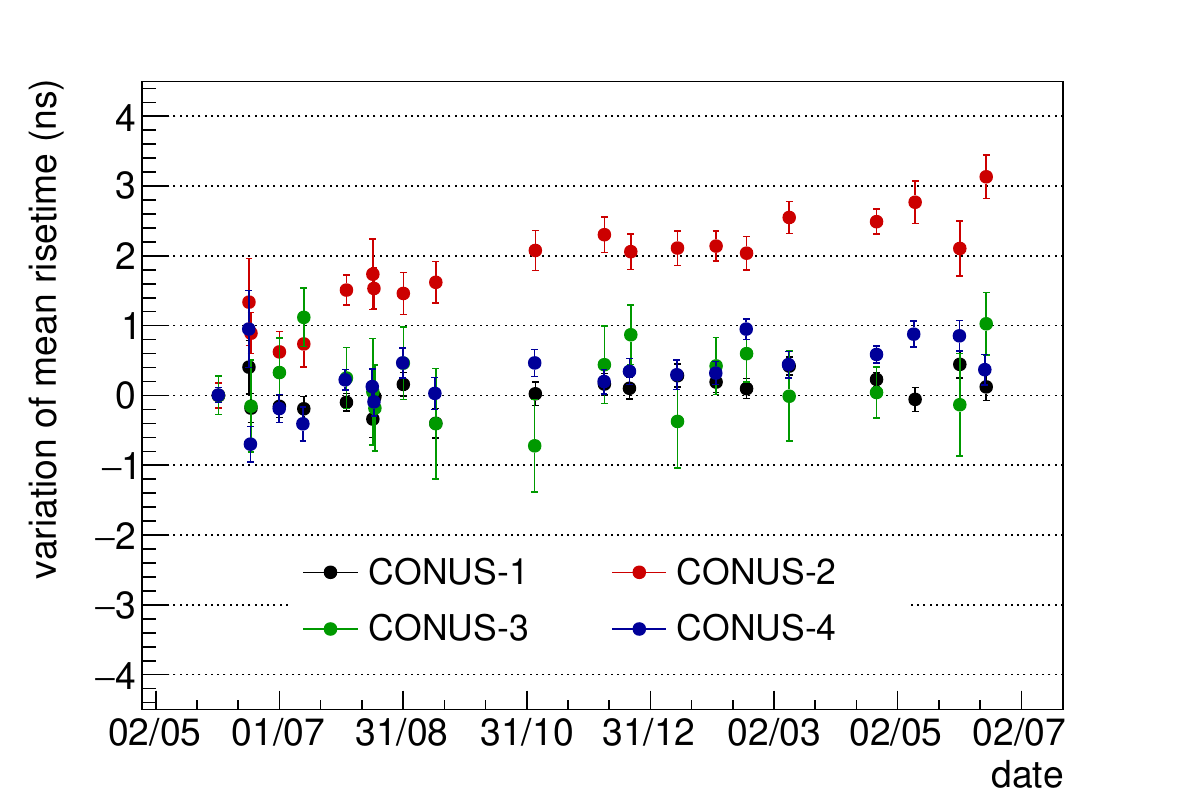}
    \caption{Deviation of the mean rise time of the bulk population over time for the four \textsc{Conus} detectors, extracted from bulk events of the $^{228}$Th calibration data with energies within (5,\,15)\keVee. The shown data cover the period from May 2021 until June 2022, corresponding to a large fraction of the \textsc{Run-5} data taking period.}
    \label{fig:stability_th228}
\end{figure}

Third, the stability of the rise time of the signals was monitored over time using the bulk event distribution of $^{228}$Th calibration data. Fig.\,\ref{fig:stability_th228} shows the mean position of the rise time distributions over a large fraction of the \textsc{Run-5} data collection period for the four \textsc{Conus} detectors. A maximal deviation of $\sim$\,3\,ns is observed for the \textsc{Conus-2} detector and is related to small instabilities of the pulse shape observed for this particular detector from $^{228}$Th measurements. This value is taken as additional source of systematic uncertainty.

\begin{figure}
    \centering
    \includegraphics[width=0.48\textwidth]{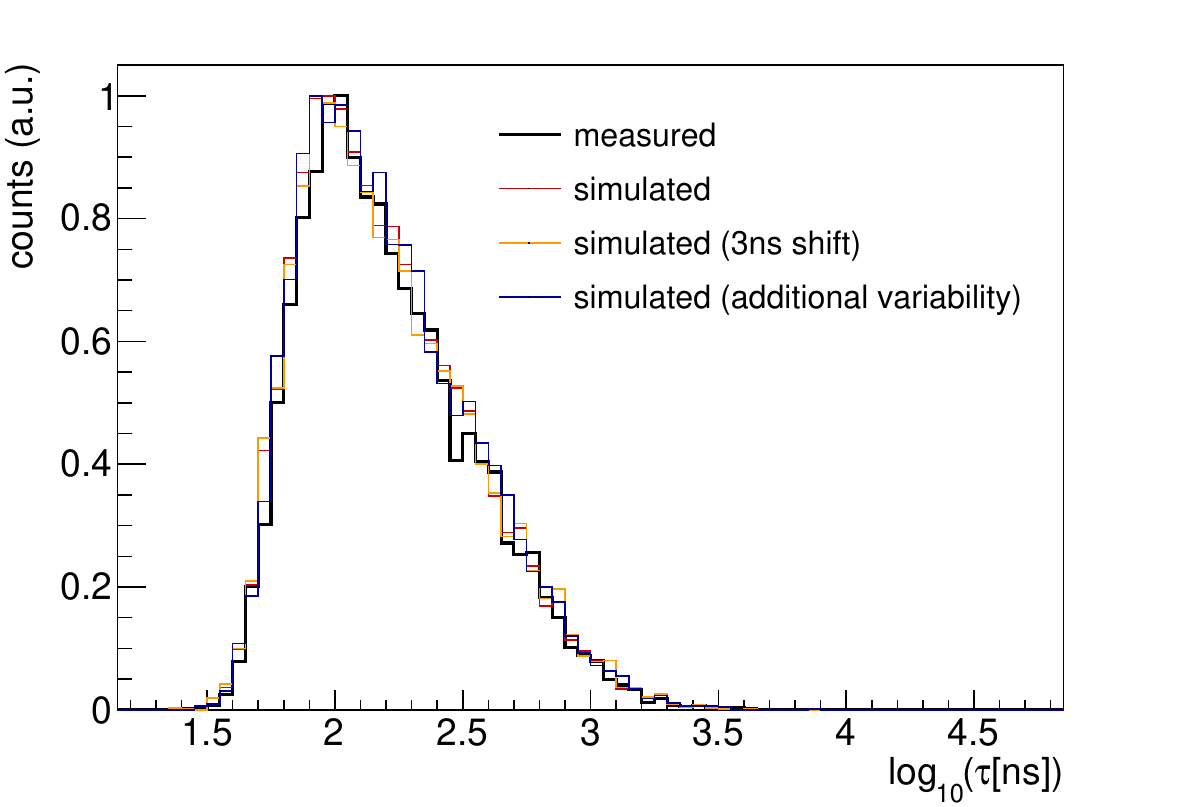}
    \caption{Measured (black) and simulated (red) rise time distributions for fast bulk events of 380\,eV$_{ee}$. The impact of a 3\,ns shift in the mean rise time of the input pulse (orange) and an additional variability in the input pulses (blue) do not affect the distribution, mostly determined by noise.}
    \label{fig:generated_noise}
\end{figure}

Fourth, an elementary simulation of event samples was developed in order to get a qualitative understanding of the broadening of the rise time distribution due to electronic noise. It also allows to investigate the impact of the small systematic uncertainties on the mean position and the width of the rise time distribution of bulk events mentioned above.
For this purpose, the theoretical input pulses mimicking bulk events are convoluted with the response function of the detector including an accurate model of the electronic noise. The noise samples are generated for each detector from their noise frequency spectrum, measured directly via dedicated baseline samples with 1\,ms acquisition windows acquired with an external arbitrary trigger. The resulting simulated rise time distributions match very well the measured distributions even at the lowest energies; an example is presented in Fig.\,\ref{fig:generated_noise}.

The impact of the systematic uncertainties on the distributions was then investigated by simulating input pulses with slightly varying rise times and by adding an artificial variability in the signals covering the small mismatches at high energy observed in the comparison with the 10.37\,keV line as well as the time variations of the rise time. As shown in Fig.\,\ref{fig:generated_noise}, such small changes do not affect the distributions which are dominated by the impact of the noise.  The systematic uncertainties discussed above can therefore safely be neglected in this energy range.

\subsection{Slow pulse rejection power}

A rise time cut can be set from the measured rise time distributions of artificial bulk events, providing a good reference for homogeneously distributed vertices in the diode. 

\begin{figure*}
		\centering
		\begin{subfigure}[c]{0.48\textwidth}
			\includegraphics[width=\textwidth]{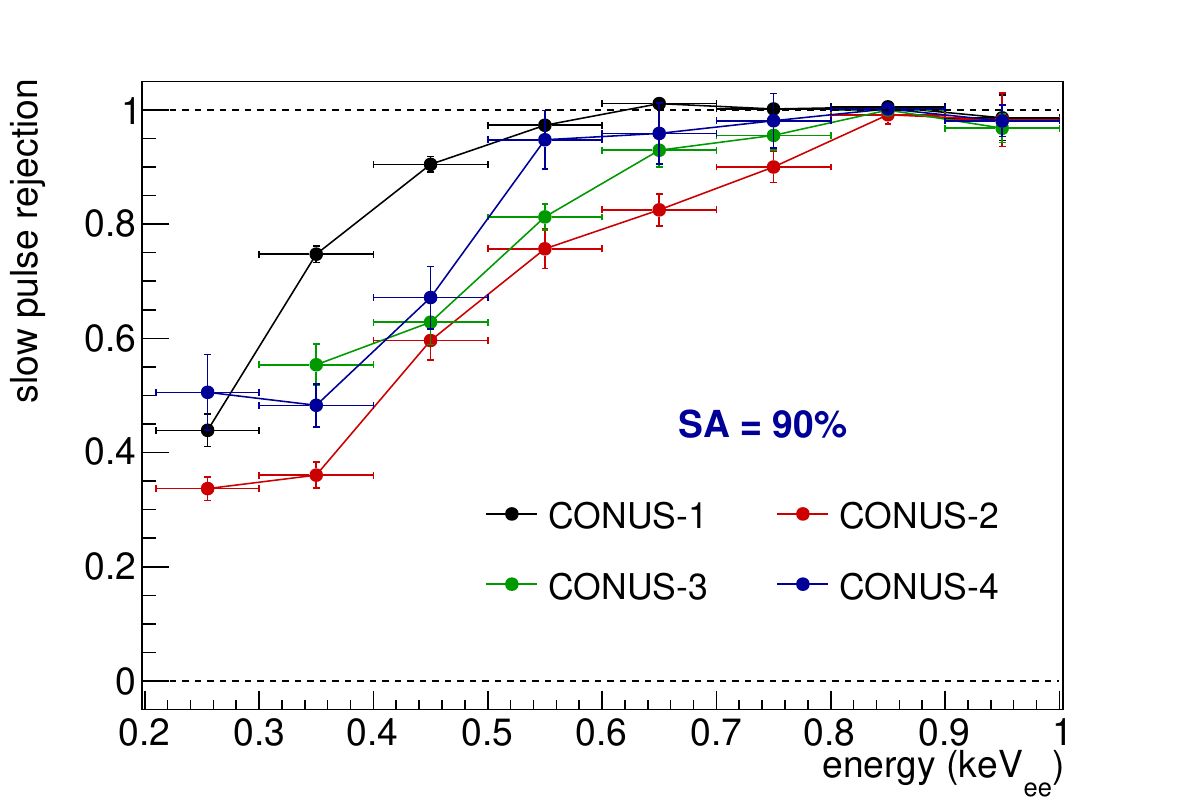}
			\caption{}\label{fig:sp_rejection_90}
		\end{subfigure}
		    \begin{subfigure}[c]{0.48\textwidth}
			\includegraphics[width=\textwidth]{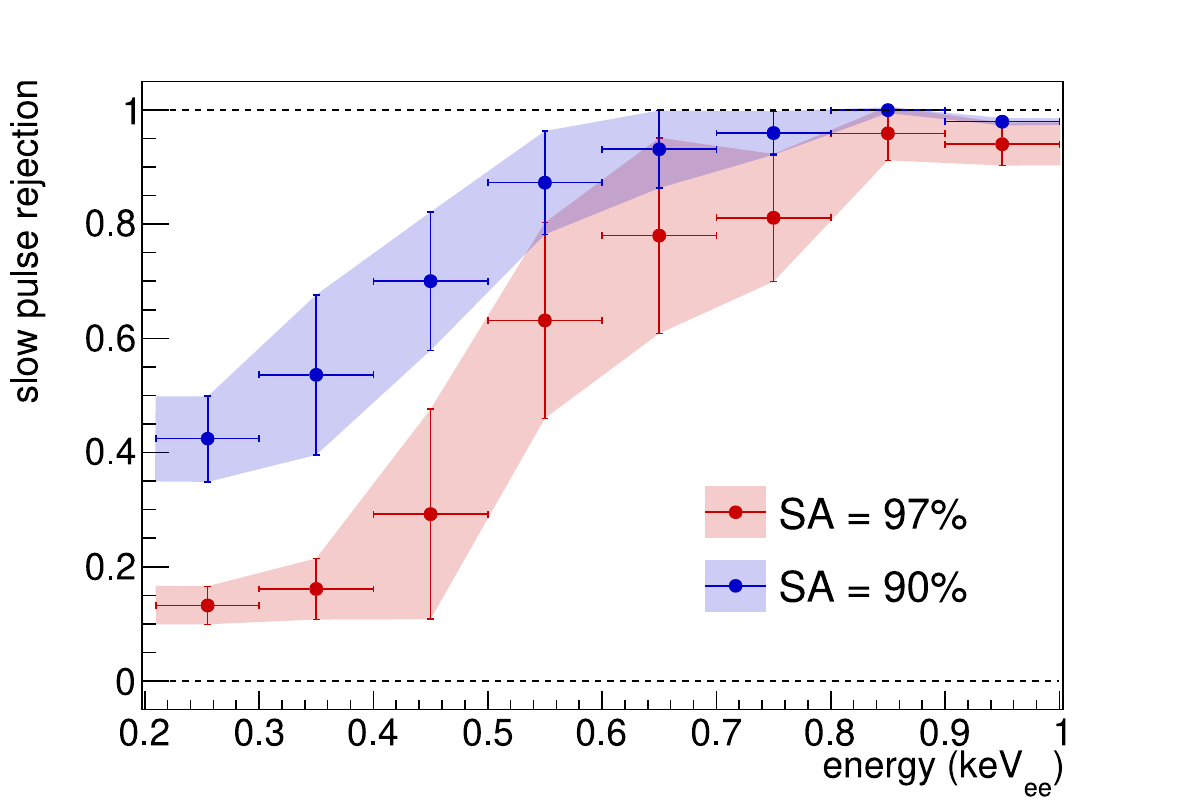}
			\caption{}\label{fig:sp_rejection_90_vs_97}
		\end{subfigure}
		\caption{Rejection efficiency of SPs as a function of the energy. The rise time distributions of the SP populations are obtained from the subtraction of the fast component from the $^{228}$Th rise time distribution.\\
  (a): Rejection efficiency for the four \textsc{Conus} detectors and a constant 90\,\% SA of FPs.\\
  (b): Average trends over the four \textsc{Conus} detectors for a SA of 97\,\% (red) and 90\,\% (blue) of SPs for comparison. The y errors bars correspond to the root-mean-square value over the detectors.\label{fig:sp_rejection} }
\end{figure*}

In order to evaluate the SP rejection capability of the PSD cut, $^{228}$Th calibration data were used as they provide a high statistics sample of SPs. The rise time distribution of the SP population only is obtained by subtraction of the FP distribution -- determined from pulser scans as described in the previous section -- from the $^{228}$Th rise time distribution. In this way, the SP rejection power can be quantified for each detector as a function of the energy for a given FP SA cut. This is illustrated in Fig.\,\ref{fig:sp_rejection_90} in the region of interest for a constant FP SA of 90\,\%. 

The uncertainties reported contain both statistic and systematic contributions. Systematic uncertainties were estimated in a conservative way by varying the integration window for the normalisation of the FP in the subtraction procedure. Note that the SP rejection numbers reported here are given for information only, helping to quantify the discrimination power of the method and allowing a direct comparison with other experiments. They are not used directly in any of the CONUS analyses since their values only affect the amount of background entering in the region of interest, which is measured directly during the reactor-off periods.
However, any drifts in time causing a time dependent SP contamination could affect the level of background entering the ROI. This is in particular crucial for the lowest energy points having the highest SP contamination. In the same way as for the FP population as shown in Fig.\,\ref{fig:generated_noise}, it was therefore cross-checked with the event generator that small rise-time drifts have a negligible impact on the SP acceptance.

Close to the $\sim$\,210\,eV$_{ee}$ energy threshold, about half of the SPs can be rejected and a perfect separation between SPs and FPs is reached between 0.7 and 0.8\,\keVee~depending on the selected detector. Variability between detectors is clearly visible, with the best rejection power obtained for the \textsc{Conus-1} detector. With the adoption of a more conservative cut -- optimized with SA of FPs of 97\,\% for the \CEvNS~analysis \cite{Stauber2022} -- a rejection of $\sim\,$15\,\% of the slow pulse component close to the energy threshold is achieved as shown in Fig.\,\ref{fig:sp_rejection_90_vs_97}.

Compared to previous similar studies with PPC HPGe detectors, this represents a notable improvement, in particular for the lowest energies: the SA reported both by TEXONO \cite{Li2014} and CDEX-1 \cite{Zhao2016} start to drop for energies $\lesssim\,$0.8\keVee~whereas for COGENT \cite{Aalseth2013}, a constant 90\,\% SA is conserved at the cost of a less strict SP rejection (less than 50\,\% rejection below 0.8\keVee). This indicates that the separation between the SP and FP populations achieved in this work is better down to lower energies. It mainly results from the very low noise level achieved for the CONUS detectors, a factor two lower than the experiments quoted above with typical pulser FWHM in the $(140-160)\,\text{eV}_{\text{ee}}$ range. As mentioned in \ref{subsection:fit}, the use of the rise-time fit was also shown to help gaining efficiency toward low energies compared to alternative simple methods, less cost-intensive but whose discriminating observables are too affected by noise for sub-keV energies \cite{Henrichs2021}.

\subsection{PSD application to CONUS background data}\label{subsection:application}

\begin{figure}
    \centering
    \includegraphics[width=0.46\textwidth]{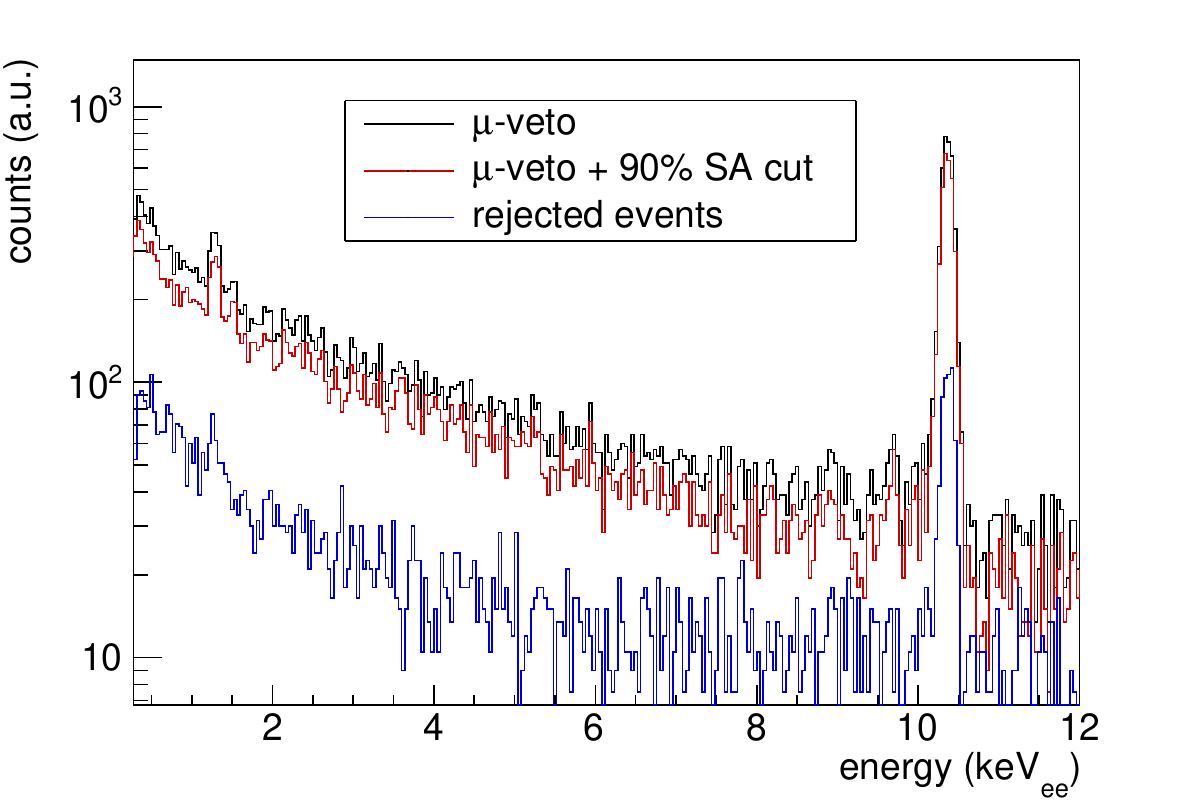}
    \caption{Background energy spectra with $\mu$-veto for \textsc{Conus-1} for \textsc{Run-5}, before (black) and after (red) the application of a PSD cut with 90\,\% SA of FPs. The rejected event spectrum is shown is blue.}
    \label{fig:background_rejection_90_95_all_dets}
\end{figure}

The pulse shape discrimination power provided by the rise time observable can be directly applied to reduce the background level of the \textsc{Conus} experiment. Indeed, due to their very large mean free path, neutrinos are expected to interact homogeneously in the entire detector and therefore mainly in the active bulk volume, whereas certain types of background radiation are expected to interact closer to the semi-active detector surface. 
Thanks to the very careful mitigation measures in the design of the detectors, the background reduction achieved for \textsc{Conus} is moderate since a large part of the background after application of the $\mu$-veto consists of FPs (see e.g. blue distributions in Fig.\,\ref{fig:psd_populations}).
The impact of a 90\,\% FP bulk-events SA cut is shown exemplary for \textsc{Conus-1} in Fig.\,\ref{fig:background_rejection_90_95_all_dets}. An overall $(15-25)\,\%$ reduction of the background level is achieved, depending on the detector and on the considered energy range.
In the \CEvNS~region of interest below 1\keVee, a background reduction of about $(15-20)\,\%$ is achieved. A more conservative selection of FP-like signals at 97\,\% still leads to a non-negligible $(5-10)\,\%$ reduction. These numbers are consistent with the MC background model expectation, predicting a contribution of SPs of about 40\,\% in this region. This number has to be coupled with the SP rejections shown in Fig.\,\ref{fig:sp_rejection_90_vs_97} to retrieve the typical background reductions found in the data.
Searches beyond the standard model carried out at higher energy ($\sim$1--10\,keV) with the \textsc{Conus} data will also benefit from this background reduction. For this region where a complete SP rejection is achieved, an overall background reduction of about 25\,\% is obtained. The gain in terms of shape of the background will also be of particular relevance: the surface event contamination is indeed expected to raise at low energy and can degenerate with most of the searched signals (\CEvNS~or beyond Standard Model physics), also having a typical rise towards low energies. Any background reduction flattening the background energy spectrum is therefore of high importance to gain sensitivity in these searches.

More generally, pulse shape studies allow to improve knowledge on the detector and can help refining and constraining the understanding of the background. An illustration of this is the study of the VFPs corresponding to interactions in the vicinity of the p+ contact, also discussed in \cite{Li2022}. At high-energy, they are distinguishable by their prompt rise time from the homogeneous bulk events as discussed in Sec.\,\ref{subsection:origin_pulses} and are responsible for the sub-component observed e.g. in Fig.\,\ref{fig:validation_10keV_ref}. Using the rise time distributions of the homogeneously distributed events from the 10.37\,keV Ge activation line, it is possible to get back to information on the weighting potential close to the p+ contact: their fraction should correspond to the volume fraction of this high $W$ region over the whole diode volume. These VFPs amount to $(8-13)\,\%$ of the bulk events, depending on the \textsc{Conus} detectors, as reported in Tab.\,\ref{tab:data_conus_detectors}. These numbers are relevant for the validation of the full pulse shape simulation, currently under development. This translates into a half-sphere of $(19.5-22.7)\,\text{mm}$ radius around the p+ contact \cite{Henrichs2021}. These relatively small fractions for \textsc{Conus} moderate the suggestion expressed in \cite{Li2022} regarding the potential of additional background reduction by discriminating this type of events. Moreover, at low energy, the two populations are not distinguishable anymore due to the overwhelming noise. Efforts in reducing the electronic noise would allow a better discrimination of these population. Moreover, a reduced p+ contact size leading to lower noise would also modify the electric potential close to the contact, which would further reduce the fraction of these VFPs. 

Using calibration sources of known energy, the rise time information is of high-relevance for the validation of the charge collection inefficiency in the transition layers as it can provide an additional and independent cross-check to the method described in \cite{Bonet2021d} using $^{241}$Am sources.

Finally, coupled to the full MC simulation of the \textsc{Conus} background, the rise time distributions are expected to provide further insights on the composition and vertices distributions of the different background components. This will allow to independently validate and complement the results already reported in \cite{Bonet2021_bkg}.

\section{Conclusion} \label{section:conclusion}
 
This article reports about the first pulse shape data analysis in the sub-keV and keV energy regime conducted in the framework of the \textsc{Conus} experiment. It exploits the capability of the used ultra-low threshold point-contact p-type high-purity germanium detectors to distinguish and discriminate surface events from bulk events by means of different signal rise time distributions.
A fitting procedure of the waveform has been selected and implemented that allows to distinguish rise times down to the detector energy thresholds of $\sim$\,210\,eV$_{ee}$.
A detailed and comprehensive understanding of the different pulse shape populations was achieved via the study of physics data (\textsc{Conus} background data and radioactive source calibration data), artificially injected data (via a pulser generator) and pulse shape simulation data.
It was verified that the pulse shapes of bulk events are homogeneous and independent of the interaction vertices except for about 10\,\% of the bulk events, which interact very close to the p+ contact and exhibit a very fast rise time.
A data-driven method was developed to generate artificially signals mimicking bulk events that can then be reliably used to characterize the bulk rise time distributions at any energy. The sub-keV$_{ee}$ region of interest was studied in detail and the growing relative influence of the electronic noise towards the noise threshold on the rise time distributions was modeled and understood.
About half of the slow pulses can be rejected at energies as low as $\sim$\,210\,eV$_{ee}$ while keeping a bulk event acceptance greater than 90\,\%. A complete slow pulses suppression is achieved above $\sim$\,800\,\,eV$_{ee}$. Thanks to these studies, a total background reduction of about $(15-25)\,\%$ is achieved for the \textsc{Conus} experiment with negligible systematic uncertainties, which agrees with our Monte Carlo background model expectation predicting $\sim\,$40\% of slow pulses in the total background of the \textsc{Conus} detectors. This additional background reduction and flattening will improve the sensitivity of the ongoing \textsc{Conus} analyses.
Beyond that, \textsc{Conus} plans further pulse shape studies to validate pulse shape simulation codes and to refine background models by providing independent information on different background components and thus reducing associated systematic uncertainties. \\

\small

\textbf{Acknowledgements}
We thank all the technical and administrative staff who helped building the experiment, in particular the MPIK workshops and Mirion Technologies (Canberra) in Lingolsheim. 
We express our gratitude to the Preussen Elektra GmbH for great support and for 
hosting the \textsc{Conus} experiment. 
The \textsc{Conus} experiment is supported financially by the Max Planck Society (MPG).

\normalsize


\bibliographystyle{bibliostyle}
\bibliography{literature}
\end{document}